\documentclass[1p]{elsarticle}
\usepackage[colorlinks]{hyperref}
\usepackage{lineno}
\modulolinenumbers[1]
\usepackage[utf8]{inputenc}
\usepackage[croatian,english]{babel}
\usepackage{amsmath}
\usepackage{amssymb}
\usepackage{enumitem}
\usepackage{arydshln} 
\usepackage[table]{xcolor}
\usepackage{stackengine}
\usepackage{makecell}

\makeatletter
\def\ps@pprintTitle{%
 \let\@oddhead\@empty
 \let\@evenhead\@empty
 \def\@oddfoot{}%
 \let\@evenfoot\@oddfoot}
\makeatother

\newcommand\xrowht[2][0]{\addstackgap[.5\dimexpr#2\relax]{\vphantom{#1}}}

\renewcommand{\vec}[1]{\boldsymbol{\mathbf{#1}}}

\journal{arXiv}

\begin{document}

\begin{frontmatter}

\title{Direct derivation of Liénard–Wiechert potentials, 
Maxwell's equations and Lorentz force  from Coulomb's law}

\author{Hrvoje Dodig\corref{ctext}}
\cortext[ctext]{Corresponding author}
\address{University of Split, 
Ruđera Boškovića 37, 21000 Split, Croatia}

\ead{hdodig@pfst.hr}

\begin{abstract}
In 19\textsuperscript{th} century Maxwell derived famous Maxwell equations
from the knowledge of three experimental physical laws: the electrostatic Coulomb's law,
Ampere's force law and Faraday's law of induction. However,  theoretical
basis for Ampere's force law and Faraday's law remains unknown to this day.
Furthermore, the Lorentz force is considered as summary of experimental
phenomena, the theoretical foundation that explains generation of this force
is still unknown.

To answer these fundamental theoretical questions,
in this paper we derive relativistically correct Liénard – Wiechert
potentials, Maxwell's equations and Lorentz force from two simple
postulates: (a) when all charges are at rest the Coulomb's force acts between the
charges, and (b) that disturbances caused by charge in motion propagate away
from the source with finite velocity. The special relativity was not used in
our derivations nor the Lorentz transformation. In effect, it was shown in
this paper that all the electrodynamic laws, including the Lorentz force, can be 
derived from  Coulomb's law and time retardation.

This was accomplished by analysis of hypothetical experiment where test charge
is at rest and where previously moving source charge stops at some time in
the past. Then the generalized Helmholtz decomposition theorem, also derived
in this paper, was applied to reformulate Coulomb's force acting at present
time as the function of positions of  source charge at previous time when the
source charge was moving. From this reformulation of Coulomb's law the
Liénard–Wiechert potentials and Maxwell's equations were derived by careful
mathematical manipulation.

In the second part of this paper, the energy conservation principle valid for
moving charges is derived  from the knowledge of electrostatic energy conservation
principle valid for stationary charges. This again was accomplished by using
generalized Helmholtz decomposition theorem. From  this dynamic energy
conservation principle  the Lorentz force is finally derived. 
\end{abstract}

\begin{keyword}
Coulomb's law \sep Liénard–Wiechert potentials \sep Maxwell equations \sep
 Lorentz force 
\end{keyword}

\end{frontmatter}


\section{Introduction}\label{sec:intro}
\noindent
In his famous \emph{Treatise}\cite{MaxwellTreatiseVol1,MaxwellTreatiseVol2} 
Maxwell derived equations of electrodynamics based on the knowledge about the
three  experimental laws known at the time: the Coulomb's law  describing  
the electric force between charges at rest; Ampere's law describing the force
between current carrying wires, and the Faraday's law of induction. Prior to
Maxwell, magnetism and electricity were regarded to as separate phenomena. It
was James Clerk Maxwell who unified these seemingly disparate phenomena into
the set of equations collectively  known today as Maxwell's equations. In
modern vector notation, the four Maxwell's equations that govern the behavior
of electromagnetic fields are written as:

\begin{align}
\label{eq_maxwell_1}
\nabla\cdot \vec{D} &= \rho \\
\label{eq_maxwell_2}
\nabla\cdot\vec{B} &= 0 \\
\label{eq_maxwell_3}
\nabla\times\vec{E} &= - \frac{\partial\vec{B}}{\partial t} \\
\label{eq_maxwell_4}
\nabla\times\vec{B} &=  \mu \vec{J} + \frac{1}{c^2} \frac{\partial\vec{E}}{\partial t}
\end{align}

\noindent
where symbol $\vec{D}$ denotes electric displacement vector, $\vec{E}$ is
electric  field vector, $\vec{B}$ is a vector called magnetic flux density,
vector $\vec{J}$ is called current density and scalar $\rho$ is the charge
density. Furthermore, there are two more important equations in
electrodynamics that relate magnetic vector potential $\vec{A}$ and scalar
potential $\phi$ to electromagnetic fields $\vec{B}$ and $\vec{E}$:

\begin{align}
\label{eq_mag_pot}
\vec{B} &=\nabla\times\vec{A} \\
\label{eq_elec_field}
\vec{E} &= -\nabla\phi -\frac{\partial \vec{A}}{\partial t}
\end{align}

\noindent
In standard electromagnetic theory, if a point charge $q_s$ is moving with 
velocity $\vec{v}_s(t)$ along arbitrary path $\vec{r}_s(t)$ the scalar
potential $\phi$ and vector potential $\vec{A}$ caused by moving charge $q_s$
are described by well known, relativistically correct, Liénard–Wiechert
potentials \cite{Lienard,Wiechert}:

\begin{align}
\label{eq_lw_pot}
\phi &=\phi(\vec{r},t)=
\frac{1}{4\pi\epsilon}\left(\frac{q_s}{\left(1-\vec{n}_s(t_r)
\cdot
\vec{\beta}_s(t_r)\right)\left|\vec{r}-\vec{r}_s(t_r)\right|}\right)
\\
\label{eq_lw_vec_pot}
\vec{A} 
&=\vec{A}(\vec{r},t)
=
\frac{\mu c}{4\pi}
\left(
\frac{q_s \vec{\beta}_s(t_r)}
{\left(1-\vec{n}_s(t_r)\cdot\vec{\beta}_s(t_r)\right)
\left|\vec{r}-\vec{r}_s(t_r)\right|}
\right)
\end{align}

\noindent
where $t_r$ is retarded time, $\vec{r}$ is the position vector of observer 
and vectors $\vec{n}_s(t_r)$ and $\vec{\beta}_s(t_r)$ are:

\begin{align}
\label{eq_vec_beta}
\vec{\beta}(t_r) &= \frac{\vec{v}_s(t_r)}{c}
\\
\label{eq_vec_n}
\vec{n}(t_r)
&=
\frac{\vec{r}-\vec{r}_s(t_r)}
{
\left|\vec{r}-\vec{r}_s(t_r)\right|
}
\end{align}

\noindent
These equations were  almost simultaneously discovered by Liénard and Wiechert
around 1900's and they represent explicit expressions for time-varying
electromagnetic fields caused by charge in arbitrary motion.  Nevertheless,
Liénard-Wiechert potentials were derived from retarded potentials, which in
turn, are derived from Maxwell equations.

Maxwell's electrodynamic equations  provide the complete description of
electromagnetic fields, however, these equations say nothing about mechanical
forces experienced by the charge moving in electromagnetic field. If the
charge $q$ is moving in electromagnetic field with velocity  $\vec{v}$ then
the force $\vec{F}$ experienced by the charge $q$ is:

\begin{equation}
\label{eq_lorentz}
\vec{F} = q\left(\vec{E}+\vec{v}\times\vec{B}\right)
\end{equation}

\noindent
The force described by equation (\ref{eq_lorentz}) is well known Lorentz
force. Discovery of this electrodynamic force is historically credited to H.A.
Lorentz \cite{Lorentz}, however, the similar expression for electromagnetic
force can be found in Maxwell's \emph{Treatise}, article 598 \cite{MaxwellTreatiseVol2}. The difference between the two is that Maxwell's electromotive force acts on moving circuits and Lorentz force acts on moving charges.

However, it is not yet explained what causes the Lorentz force, Ampere's force law and Faraday's law. Maxwell derived his expression for electromotive force along moving circuit from the knowledge of experimental Faraday's law. Later, Lorentz extended Maxwell's reasoning to discover the force acting on charges moving  in electromagnetic field \cite{Lorentz}. Nevertheless,  it would be impossible for Lorentz to derive his force law without the prior knowledge of Maxwell equations \cite{Darringol1994}.

Nowadays, the Lorentz force  ($q\vec{v}\times\vec{B}$ term) is commonly viewed as an effect of Einstein's special relativity. For example,  an observer co-moving with source charge would not measure any magnetic field, while on the other hand, the stationary observer would measure the magnetic field caused by moving source charge. However, in this work, we demonstrate that the special relativity is not needed to derive the Lorentz force and Maxwell equations. In fact, we derive Maxwell's equations and Lorentz force from more fundamental principles: the Coulomb's law and time retardation.

There is another reason why the idea to derive Maxwell's equations and Lorentz force from Coulomb's law may seem plausible. Because of mathematical similarity between
Coulomb's law and Newton's law of gravity many researchers thought that if Maxwell's equations and Lorentz force could be derived from Coulomb's law that this would be helpful in understanding of gravity. These two inverse-square physical laws are written:

\begin{align}
\label{eq_coulomb}
\vec{F}_C &= \frac{q_1 q_2}{4\pi\epsilon} \frac{\vec{r}_1-\vec{r}_2}
{\left|\vec{r}_1-\vec{r}_2\right|^3}
\\
\label{eq_newton}
\vec{F}_G &= -G m_1 m_2\frac{\vec{r}_1-\vec{r}_2}
{\left|\vec{r}_1-\vec{r}_2\right|^3}
\end{align}

\noindent
The expressions for Coulomb's force and Newton's gravitational force are
indeed similar, however, these two forces   significantly differ in physical
nature. The latter force is always attractive while the former can be either
attractive or repulsive. Nevertheless, a number of researchers attempted to
derive Maxwell's equations from Coulomb's law, and most of these attempts rely
on Lorentz transformation of space-time coordinates between the rest frame of
the moving charge and laboratory frame.

The first hint that Maxwell's equations could be derived from Coulomb's law
and Lorentz transformation can be found in Einstein's original 1905 paper on
special relativity \cite{Einstein1905}.  Einstein suggested that the Lorentz
force term ($\vec{v}\times\vec{B}$) is to be attributed to Lorentz
transformation of the electrostatic field from the rest frame of  moving
charge to the laboratory frame where the charge has constant velocity. Later,
in 1912, Leigh Page derived Faraday's law and Ampere's law  from Coulomb's law
using Lorentz transformation \cite{Page1912}. Frisch and Willets discussed the
derivation of Lorentz force from Coulomb's law using relativistic
transformation of force \cite{Frisch1956}. Similar route to derivation of
Maxwell's equations and Lorentz force from Coulomb's law was taken by Elliott
in 1966 \cite{Elliott1966}. Kobe in 1986 derives Maxwell's equations as the
generalization of Coulomb's law using special relativity \cite{Kobe1986}.
Lorrain and Corson derive Lorentz force from Coulomb's law, again, by using
Lorentz transformation and special relativity\cite{Lorain1988}.  Field in 2006
derives Lorentz force and magnetic field starting from Coulomb's law by
relating the electric field  to electrostatic potential in a manner consistent
with special relativity \cite{Field2006}. The most recent attempt comes from
Singal \cite{Singal2011} who attempted to derive electromagnetic fields of
accelerated charge from Coulomb's law and Lorentz transformation.

All of the  mentioned attempts  have in common that they attempt to derive
Maxwell equations from Coulomb's law by exploiting Lorentz transformation or
Einstein's special theory of relativity. However, historically the Lorentz
transformation was derived from Maxwell's equations \cite{Lorentz1904}, thus,
the attempt to  to derive Maxwell's equations using Lorentz transformation 
seems to involve circular reasoning \cite{Rothwell2018}.
The strongest criticism came from Jackson who pointed out that it should be
immediately obvious that, without additional assumptions, it is impossible to
derive Maxwell's equations from Coulomb's law using theory of special
relativitiy \cite{Jackson1975}. Schwartz  addresses these additional
assumptions and starting from Gauss' law of electrostatics and by exploiting
the Lorentz invariance and properties of Lorentz transformation he derives the
Maxwell's equations \cite{Schwartz1987}.

In addition to the criticism above, we point out that the derivations of
Maxwell's equations from Coulomb's law using Lorentz transformation should
only  be considered valid for the special case of the charge moving along the
straight line with constant velocity. This is because the Lorentz
transformation is derived under the assumption that electron moves with
constant velocity along straight line \cite{Lorentz1904}. For example, if the
particle moves with uniform acceleration along straight line the
transformation of coordinates between the rest frame of the particle and the
laboratory frame takes the different mathematical form than that of the
Lorentz transformation \cite{Rosen1962}. If the particle is in uniform
circular motion yet another coordinate transformation from the rest frame to
laboratory frame, called Franklin transform, is valid \cite{Franklin1922}.
None of the above cited papers consider the fact that Lorentz transformation
is no longer valid when the charge is not moving along straight line with
constant velocity.

\begin{figure}[!h]
\centerline{\includegraphics[width=.95\textwidth]{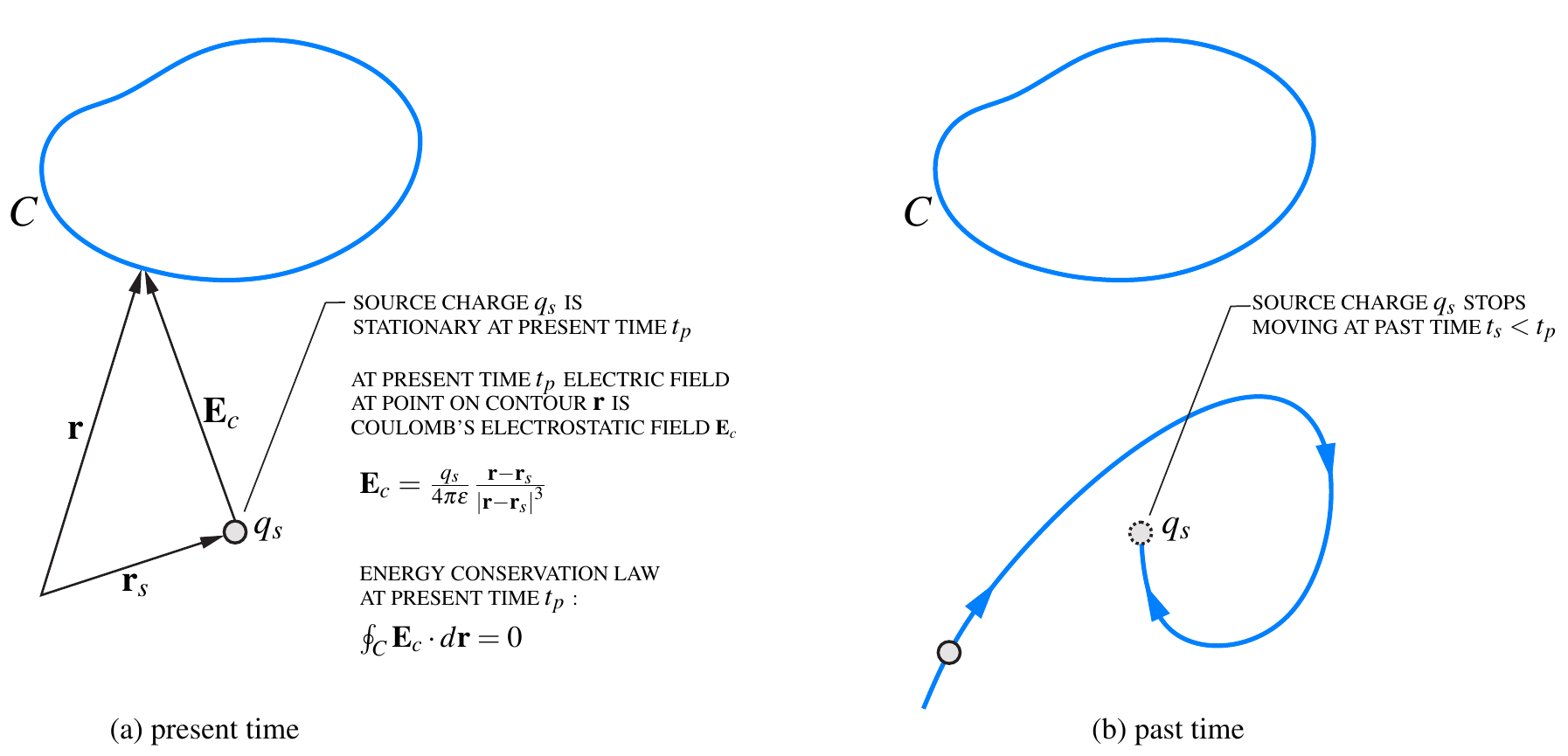}}
\caption{In (a) the source charge $q_s$ is at rest at present time $t_p$. 
Each point on closed contour $C$ is affected by Coulomb's electrostatic field
$\vec{E}_c$. Energy conservation principle at present time is 
$\oint_{C}{\vec{E}_c\cdot d\vec{r}}=0$. In (b) the source charge $q_s$  is
moving along arbitrary path and it stops at past time $t_s<t_p$. Dynamic
energy conservation principle valid in the past is assumed to be unknown
when source charge was moving.}
\label{fig1}
\end{figure}

To circumvent problems with special relativity and Lorentz transformation we
take entirely different approach to derive Liénard–Wiechert potentials and
Maxwell's equations from Coulomb law. We start our derivation from the
analysis of the following hypothetical experiment: consider two charges at
rest at present time, one called the test charge, and the other called the
source charge. The source charge was moving in the past but it is at rest at 
present time. Because both  charges are at rest at present  the force
acting on test charge at present time is the Coulomb's force. 

However, in the past when the source charge was moving, we assume that the force acting on test charge was not the Coulomb's force. To discover the mathematical form of this "unknown" electrodynamic force acting in the past
from the knowledge of known electrostatic force (Coulomb's law) acting at present time the generalized Helmholtz decomposition theorem was applied. This
theorem, derived in  \ref{sec:apx_gen_helmholtz}, allowed us to relate
Coulomb's force acting at present time to the positions of source charge at past
time. From here, Liénard–Wiechert potentials and Maxwell's equations were
derived by careful mathematical manipulation. 

It should be emphasized that we did not resort to theory of special relativity
nor to Lorentz transformation in our derivation of Maxwell's equations. Not
less importantly, the presented derivation of Maxwell's equations from
Coulomb's law is valid for charges in arbitrary motion. In effect, we may say
that  more general physical law (Maxwell's equations) acting at
past time is derived from  the knowledge of limited physical law (Coulomb's law) acting
at  present time.

However, from Maxwell's equations, it is very difficult, if not entirely impossible, to derive 
the Lorentz force without resorting to some form of energy conservation law.
As shown in Fig. \ref{fig1}a, at present time, the single stationary charge creates Coulomb's electrostatic field. Known energy conservation law valid at present time states that contour integral of Coulomb's field along closed contour $C$ is equal to zero.

But, this electrostatic energy conservation law valid at present is not necessarily valid in the past when the source charge was moving. Thus, in the second part of this paper we derive this "unknown" dynamic energy
conservation principle valid in the past from the knowledge of electrostatic energy conservation principle
valid at present time. This was again achieved
by the careful application of generalized Helmholtz decomposition theorem which allowed us to transform electrostatic energy conservation law valid at present to dynamic energy conservation law valid in the past.

This dynamic
energy conservation law states that the work of non-conservative force along
closed contour is equal to the time derivative of the flux of certain vector
field through the surface bounded by this closed contour.  From this dynamic energy conservation law the
Lorentz force was finally derived.

\section{Generalized Helmholtz  decomposition theorem}
Because generalized Helmholtz decomposition theorem  is central for deriving Maxwell equations and Lorentz force from Coulomb's law, in this section, we briefly present this important theorem while the derivation itself is moved to \ref{sec:apx_gen_helmholtz}.
There have been several previous attempts in the literature to generalize
classical Helmholtz decomposition theorem to time dependent vector fields
\cite{Hauser1970,Kapuscik1985,Davis2006}. However, in none of the cited articles the
Helmholtz theorem for functions of space and time is presented in the mathematical
form usable for the mathematical developments described in this paper. This is
probably caused by difficulties in stating such a theorem and this was clearly
stated in \cite{Kapuscik1985}:  "There does not exist
any simple generalization of this theorem for time-dependent
vector fields". 

However, we show that there indeed exists the simple
generalization of Helmholtz decomposition theorem for time-dependent vector
fields and  that it can  derived from  time-dependent inhomogeneous wave equation. 
To improve the clarity of this paper, the complete derivation of Helmholtz
decomposition theorem for functions of space and time is moved to 
\ref{sec:apx_gen_helmholtz}, subsection \ref{sec_gen_helmholtz}.  As it was shown in \ref{sec:apx_gen_helmholtz}, the
generalization of Helmholtz decomposition theorem for the vector function of
space and time $\vec{F}(\vec{r},t)$ can be written as:

\begin{align}
\label{eq_ext_Helmholtz}
\vec{F}(\vec{r},t)
=
&-\nabla \int_{\mathbb{R}}{
dt'
\int_{\mathbb{R}^3}{
\bigg(
\nabla'\cdot\vec{F}(\vec{r}',t')
\bigg)
G(\vec{r},t;\vec{r}',t')
dV'
}
}
\\
&+
\frac{1}{c^2}
\frac{\partial}{\partial t}
 \int_{\mathbb{R}}{
dt'
\int_{\mathbb{R}^3}{
\left(
\frac{\partial}{\partial t'}\vec{F}(\vec{r}',t')
\right)
G(\vec{r},t;\vec{r}',t')
dV'
}
}
\nonumber
\\
&+\nabla\times \int_{\mathbb{R}}{
dt'
\int_{\mathbb{R}^3}{
\bigg(
\nabla'\times\vec{F}(\vec{r}',t')
\bigg)
G(\vec{r},t;\vec{r}',t')
dV'
}
}
\nonumber
\end{align}

\noindent
where scalar function $G(\vec{r},t;\vec{r}',t')$ is the fundamental solution
of time-dependent inhomogeneous wave equation given as:

\begin{equation}
\label{eq_inhom_wave}
\nabla^2 G(\vec{r},t;\vec{r}',t')
-
\frac{1}{c^2}
\frac{\partial^2}{\partial t^2}
G(\vec{r},t;\vec{r}',t')
=
-\delta(\vec{r}-\vec{r}')\delta(t-t')
\end{equation}

\noindent
In the equation above, 
$\delta(\vec{r}-\vec{r}')=\delta(x-x')\delta(y-y')\delta(z-z')$ is 3D Dirac
delta function, and $\delta(t-t')$ is Dirac delta function in one dimension.
Fundamental solution $G(\vec{r},t;\vec{r}',t')$, sometimes called Green's
function, represents the retarded in time solution of the inhomogeneous time
dependent wave equation and it can be written as:

\begin{equation}
\label{eq_fund_wave}
G(\vec{r},t;\vec{r}',t')
=
\frac{
\delta\left( t'-t+\frac{\left|\vec{r}-\vec{r}'\right|}{c}\right)
}{
4\pi \left|\vec{r}-\vec{r}'\right|
}
\end{equation}

\noindent
where position vector $\vec{r}'$ is the location of the source at time $t'$.

From equation (\ref{eq_ext_Helmholtz}) it is evident that the Helmholtz
decomposition theorem for functions of space and time can be regarded to as a
mathematical tool that allows us to rewrite any  vector function that is
function of present time $t$ and of present position $\vec{r}$ as   vector
function of previous time $t'$ and of previous position $\vec{r'}$.
Furthermore, the generalized Helmholtz decomposition theorem
(\ref{eq_ext_Helmholtz}) comes with additional limitation that it is  valid if
vector function $\vec{F}(\vec{r}',t')$ approaches zero faster than 
$1/\left|\vec{r}-\vec{r}'\right|$ as 
$\left|\vec{r}-\vec{r}'\right|\rightarrow\infty$.

 Very similar theorem was presented in article written by Heras
 \cite{Heras2016}; the difference is that in Heras' article the time integrals
 in equation (\ref{eq_ext_Helmholtz}) were a priori evaluated at retarded time
 $t'=t-\left|\vec{r}-\vec{r}'\right|/c$. As such, the generalized Helmholtz
 theorem presented in \cite{Heras2016} is not suitable for the derivation of
 Maxwell equations and Lorentz force from the Coulomb's law. Reason for this,
 as it will become evident later in this paper, is that if we immediately
 evaluate the time integrals  in equation (\ref{eq_ext_Helmholtz}) the
 important information is lost from the equation.

\section{Derivation of Maxwell Equations from Coulomb's law}\label{sec:maxwell_deriv}
\begin{figure}[!t]
\centerline{\includegraphics[width=.8\textwidth]{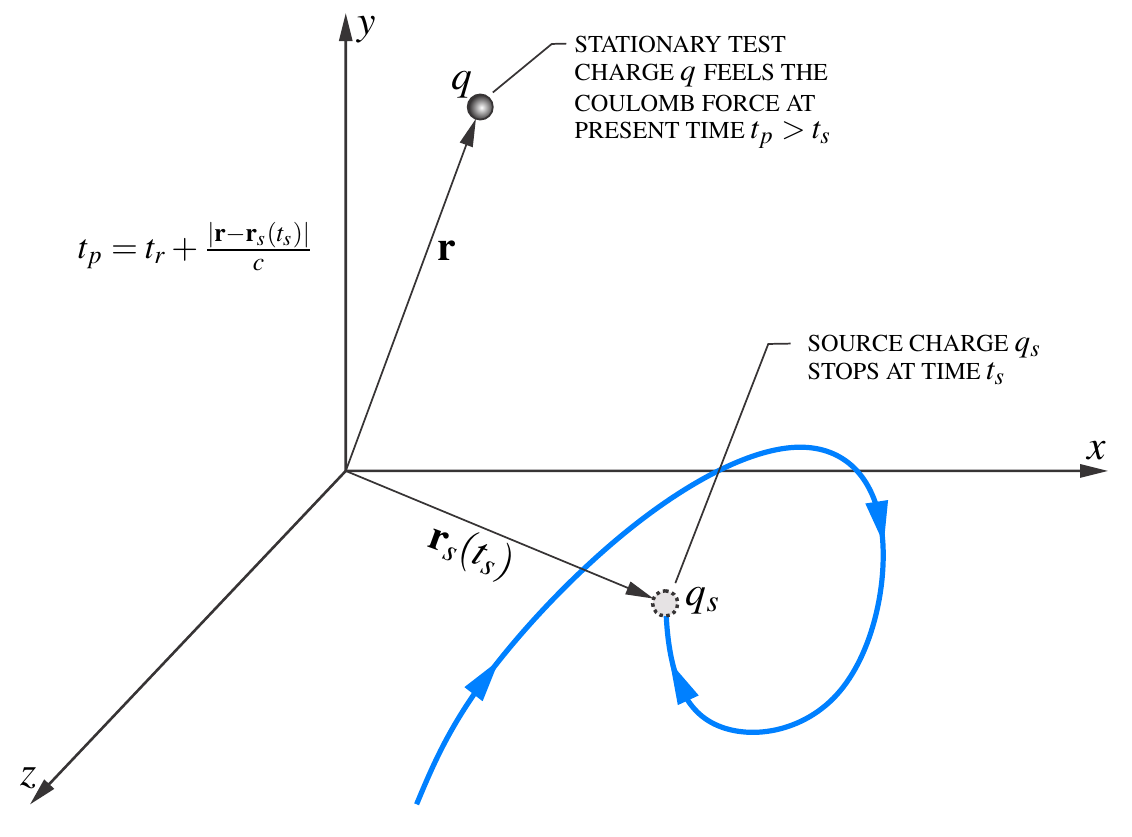}}
\caption{Source charge $q_s$ is moving along arbitrary trajectory 
$\vec{r}_s(t)$ and stops at time $t_s$. Because $q_s$ stops moving at
past time $t_s$, at present time $t_p$, the stationary test charge $q$
experiences electrostatic Coulomb force.}
\label{fig2}
\end{figure}

In this section we derive Maxwell equations from Coulomb's law using 
generalized Helmholtz decomposition theorem  represented by equation
(\ref{eq_ext_Helmholtz}). To begin the discussion,  we consider hypothetical
 experiment shown in Fig. \ref{fig2}, where source charge $q_s$ is
moving along trajectory $\vec{r}_s(t)$ and  it stops at some past time $t_s$.
The test charge $q$ is stationary at all times. We assume that the
disturbances caused by moving source charge propagate outwardly from the
source charge  with finite velocity $c$. These disturbances
originating from the source charge at past time manifest itself as the force acting on stationary test charge at present
time.  This means that there is
a time delay $\Delta t$ between the past time  $t_s$ when the source charge
has stopped and the present time $t_p$ when this disturbance has propagated to
the test charge:

\begin{equation}
\label{eq_time_retardation}
\Delta t = t_p-t_s=\frac{\left|\vec{r}-\vec{r}_s(t_s)\right|}{c}
\end{equation}

\noindent
At precise moment  in time $t_p$, that we call the present time, the force
acting on stationary test charge $q$ is the Coulomb's force because  source
charge and test charge are both at rest, and because the effect of source
charge stopping at  past time $t_s$ had enough time to propagate to test
charge. The Coulomb's force $\vec{F}_c$ experienced by the test charge $q$ 
at present time $t_p$ can  be expressed by the following equation:

\begin{equation}
\label{eq_coulomb_d1}
\vec{F}_c(\vec{r},t_p)
=
\frac{q q_s}{4 \pi \epsilon}
\frac{\vec{r}-\vec{r}_s(t_s)}
{\left|\vec{r}-\vec{r}_s(t_s)\right|^3}
\hspace{3em} t_p= t_s + \Delta t
\end{equation}

\noindent
Let us now consider the time  $t$ just one brief moment before 
the stopping time $t_s$:

\begin{equation}
\label{eq_coulomb_d1_1}
t = t_s - \delta t
\end{equation}

\noindent
where $\delta t\rightarrow 0$ is very small time interval. This time interval
$\delta t$ is so small that we might even call it infinitesimally small. Then at
the moment in time infinitesimally before the present time $t_p$ the force
felt by test charge $q$ is still the Coulomb's force if 
$\delta t\rightarrow 0$. Using these considerations, we can now rewrite equation (\ref{eq_coulomb_d1})  as:

\begin{equation}
\label{eq_coulomb_d1_2}
\vec{F}_c(\vec{r},t_p-\delta t)
=
\frac{q q_s}{4 \pi \epsilon}
\frac{\vec{r}-\vec{r}_s(t)}
{\left|\vec{r}-\vec{r}_s(t)\right|^3}
\hspace{3em} t= t_s - \delta t;
\hspace{.6em} \delta t\rightarrow 0
\end{equation}

\noindent
Note that equation (\ref{eq_coulomb_d1_2}) is equivalent to equation (\ref{eq_coulomb_d1}) when $\delta t\rightarrow 0$.
The reason why we have written the Coulomb's law this way is to permit slight
variation of time before stopping time $t_s$ so that we can exploit
generalized Helmholtz decomposition theorem in order to derive Maxwell's equations
from Coulomb's law. Had we not done this then the source charge position
vector $\vec{r}_s(t_s)$ would simply be the constant vector and generalized
Helmholtz decomposition could not be used. 

Because the right hand side of equation (\ref{eq_coulomb_d1_2}) is now
the function of time $t$ and position   $\vec{r}$  we are allowed
to use the generalized Helmholtz decomposition theorem to rewrite the right
hand side of equation (\ref{eq_coulomb_d1_2}). This is  because generalized Helmholtz
decomposition theorem states that \emph{any} vector function of time $t$ and
position $\vec{r}$ can be decomposed as described by this theorem if that
function meets certain criteria. Thus, using generalized Helmholtz decomposition
theorem  we can rewrite the right
hand side of equation (\ref{eq_coulomb_d1_2}) as:

\begin{eqnarray}
\label{eq_coulomb_start}
\lefteqn{
\vec{F}_c(\vec{r},t_p-\delta t)
=
\frac{q q_s}{4 \pi \epsilon}
\frac{\vec{r}-\vec{r}_s(t)}
{\left|\vec{r}-\vec{r}_s(t)\right|^3}
=
}
\\
&\hspace{5em}=&
-\nabla \int_{\mathbb{R}}{
dt'
\int_{\mathbb{R}^3}{
\bigg(
\nabla'\cdot \frac{q q_s}{4 \pi \epsilon}
\frac{\vec{r}'-\vec{r}_s(t')}
{\left|\vec{r}'-\vec{r}_s(t')\right|^3}
\bigg)
G(\vec{r},t;\vec{r}',t')
dV'
}
}
\nonumber
\\
&&+
\frac{1}{c^2}
\frac{\partial}{\partial t}
 \int_{\mathbb{R}}{
dt'
\int_{\mathbb{R}^3}{
\left(
\frac{\partial}{\partial t'}\frac{q q_s}{4 \pi \epsilon}
\frac{\vec{r}'-\vec{r}_s(t')}
{\left|\vec{r}'-\vec{r}_s(t')\right|^3}
\right)
G(\vec{r},t;\vec{r}',t')
dV'
}
}
\nonumber
\\
&&+\nabla\times \int_{\mathbb{R}}{
dt'
\int_{\mathbb{R}^3}{
\bigg(
\nabla'\times\frac{q q_s}{4 \pi \epsilon}
\frac{\vec{r}'-\vec{r}_s(t')}
{\left|\vec{r}'-\vec{r}_s(t')\right|^3}
\bigg)
G(\vec{r},t;\vec{r}',t')
dV'
}
}
\nonumber
\end{eqnarray}

\noindent
To clarify the notation in equation above note that
$dV'=dx'dy'dz'$ represents the differential volume element of an infinite
volume $\mathbb{R}^3$. As defined in   \ref{sec:apx_gen_helmholtz},
subsection \ref{sec:apx_identities_prelim}, the primed position vector $\vec{r}'$ 
is written in Cartesian coordinate system as:

\begin{equation}
\label{eq_primed_r}
\vec{r}' = x'\vec{\hat{x}} + y'\vec{\hat{y}} + z'\vec{\hat{z}}
\end{equation}

\noindent
where variables $x',y',z'\in\mathbb{R}$.
Vectors $\vec{\hat{x}}$, $\vec{\hat{y}}$ and $\vec{\hat{z}}$ are orthogonal
Cartesian unit basis vectors. Furthermore, in Cartesian coordinates,
the primed del operator $\nabla'$ that appears in equation
(\ref{eq_coulomb_start}) is defined as:

\begin{equation}
\label{eq_primed_del}
\nabla' = \vec{\hat{x}}\frac{\partial}{\partial x'} +
          \vec{\hat{y}}\frac{\partial}{\partial y'} +
          \vec{\hat{z}}\frac{\partial}{\partial z'}
\end{equation}

\noindent
From the definition above, it follows that primed  del operator $\nabla'$ acts
only on functions of variables $x',y',z'$, and consequently, on functions of
primed position vector 
$\vec{r}'=x'\vec{\hat{x}}+y'\vec{\hat{y}}+z'\vec{\hat{z}}$. It does not act 
on functions of position vector of source charge $\vec{r}_s(t')$ because this
position vector is function of variable $t'$. Using these definitions we can
write the following simple relations:

\begin{align}
\label{eq_utility_1}
\frac{\vec{r}'-\vec{r}_s(t')}
{\left|\vec{r}'-\vec{r}_s(t')\right|^3}
&= -\nabla' \frac{1}{\left|\vec{r}'-\vec{r}_s(t')\right|}
\\
\label{eq_utility_2}
\nabla'\cdot
\frac{1}{4\pi}
\frac{\vec{r}'-\vec{r}_s(t')}
{\left|\vec{r}'-\vec{r}_s(t')\right|^3}
&=
-\nabla'^2
\frac{1}{4\pi}
 \frac{1}{\left|\vec{r}'-\vec{r}_s(t')\right|}
=
\delta\left(\vec{r'}-\vec{r}_s(t')\right)
\\
\label{eq_utility_3}
\nabla' \times
\frac{\vec{r}'-\vec{r}_s(t')}
{\left|\vec{r}'-\vec{r}_s(t')\right|^3}
&=
 -\nabla' \times \nabla'\frac{1}{\left|\vec{r}'-\vec{r}_s(t')\right|}
 =
 0
\end{align}

\noindent
where $\delta\left(\vec{r'}-\vec{r}_s(t')\right)$ is 3D Dirac's delta
function. Inserting equations (\ref{eq_utility_2}) and (\ref{eq_utility_3})
into equation (\ref{eq_coulomb_start}), and eliminating charge $q$ from the
equation, yields the following relation:

\begin{align}
\label{eq_coulomb_d2}
\frac{q_s}{4 \pi \epsilon}
\frac{\vec{r}-\vec{r}_s(t)}
{\left|\vec{r}-\vec{r}_s(t)\right|^3}
=
&-\nabla \int_{\mathbb{R}}{
dt'
\int_{\mathbb{R}^3}{
\frac{q_s}{\epsilon}
\delta\left(\vec{r'}-\vec{r}_s(t')\right)
G(\vec{r},t;\vec{r}',t')
dV'
}
}
\\
&+
\frac{1}{c^2}
\frac{\partial}{\partial t}
\int_{\mathbb{R}}{
dt'
\int_{\mathbb{R}^3}{
\left(
\frac{\partial}{\partial t'}\frac{q_s}{4 \pi \epsilon}
\frac{\vec{r}'-\vec{r}_s(t')}
{\left|\vec{r}'-\vec{r}_s(t')\right|^3}
\right)
G(\vec{r},t;\vec{r}',t')
dV'
}
}
\nonumber
\end{align}

\noindent
In \ref{sec:apx_identities}, subsection \ref{sec:apx_identities_d1}, we
have shown that the time derivative that appears in the second right hand 
side integral of equation (\ref{eq_coulomb_d2}) can be written as: 

\begin{equation}
\label{eq_coulomb_d3}
\frac{\partial}{\partial t'}\frac{q_s}{4 \pi \epsilon}
\frac{\vec{r}'-\vec{r}_s(t')}
{\left|\vec{r}'-\vec{r}_s(t')\right|^3}
=
\frac{q_s}{4 \pi \epsilon}
\nabla'
\times
\nabla'
\times
\frac{\vec{v}_s(t')}{\left|\vec{r}'-\vec{r}_s(t')\right|}
- \frac{q_s}{\epsilon} \vec{v}_s(t') \delta\left(\vec{r}'-\vec{r}_s(t')\right)
\end{equation}

\noindent
\noindent
where $\vec{v}_s(t')$ is the velocity of the source charge $q_s$ at time $t'$:

\begin{equation}
\label{eq_velocity}
\vec{v}_s(t')
=
\frac{\partial \vec{r}_s(t')}{\partial t'}
\end{equation}

\noindent
By inserting equation (\ref{eq_coulomb_d3}) into equation
(\ref{eq_coulomb_d2}) it is obtained that:

\begin{align}
\label{eq_coulomb_d9}
\frac{q_s}{4 \pi \epsilon}
\frac{\vec{r}-\vec{r}_s(t)}
{\left|\vec{r}-\vec{r}_s(t)\right|^3}
=
&-\nabla \int_{\mathbb{R}}{
dt'
\int_{\mathbb{R}^3}{
\frac{q_s}{\epsilon}
\delta\left(\vec{r'}-\vec{r}_s(t')\right)
G(\vec{r},t;\vec{r}',t')
dV'
}
}
\\
&-
\frac{1}{c^2}
\frac{\partial}{\partial t}
\int_{\mathbb{R}}{
dt'
\int_{\mathbb{R}^3}{
\frac{q_s}{\epsilon} \vec{v}_s(t') \delta\left(\vec{r}'-\vec{r}_s(t')\right)
G(\vec{r},t;\vec{r}',t')
dV'
}
}
\nonumber
\\
&+
\frac{1}{c^2}
\frac{\partial}{\partial t}
\int_{\mathbb{R}}{
dt'
\int_{\mathbb{R}^3}{
\frac{q_s}{4 \pi \epsilon}
\left[
\nabla'
\times
\nabla'
\times
\frac{\vec{v}_s(t')}{\left|\vec{r}'-\vec{r}_s(t')\right|}
\right]
G(\vec{r},t;\vec{r}',t')
dV'
}
}
\nonumber
\end{align}

\noindent
We now make use of the following identity, also derived in Appendix
\ref{sec:apx_identities}, subsection \ref{sec:apx_identities_d2}, to rewrite the
last right hand side term of equation (\ref{eq_coulomb_d9}) as:

\begin{eqnarray}
\label{eq_coulomb_d10}
\lefteqn{
\int_{\mathbb{R}}{
dt'
\int_{\mathbb{R}^3}{
\frac{q_s}{4 \pi \epsilon}
\left[
\nabla'
\times 
\nabla'
\times
\frac{\vec{v}_s(t')}{\left|\vec{r}'-\vec{r}_s(t')\right|}
\right]
G(\vec{r},t;\vec{r}',t')
dV'
}
}
=
}
\\
&=&
\nabla
\times
\nabla
\times
\int_{\mathbb{R}}{
dt'
\int_{\mathbb{R}^3}{
\frac{q_s}{4 \pi \epsilon}
\frac{\vec{v}_s(t')}{\left|\vec{r}'-\vec{r}_s(t')\right|}
G(\vec{r},t;\vec{r}',t')
dV'
}
}
\nonumber
\end{eqnarray}

\noindent
Replacing the last right hand side integral in equation (\ref{eq_coulomb_d9})
with equation (\ref{eq_coulomb_d10}) and differentiating the resulting
equation with respect to time $t$ yields:

\begin{align}
\label{eq_coulomb_d11}
\frac{q_s}{4 \pi \epsilon}
\frac{\partial}{\partial t}
\frac{\vec{r}-\vec{r}_s(t)}
{\left|\vec{r}-\vec{r}_s(t)\right|^3}
=
&-\frac{\partial}{\partial t} \nabla \int_{\mathbb{R}}{
dt'
\int_{\mathbb{R}^3}{
\frac{q_s}{\epsilon}
\delta\left(\vec{r'}-\vec{r}_s(t')\right)
G(\vec{r},t;\vec{r}',t')
dV'
}
}
\\
&-
\frac{1}{c^2}
\frac{\partial^2}{\partial t^2}
\int_{\mathbb{R}}{
dt'
\int_{\mathbb{R}^3}{
\frac{q_s}{\epsilon} \vec{v}_s(t') \delta\left(\vec{r}'-\vec{r}_s(t')\right)
G(\vec{r},t;\vec{r}',t')
dV'
}
}
\nonumber
\\
&+
\frac{1}{c^2}
\frac{\partial^2}{\partial t^2}
\nabla\times\nabla\times
\int_{\mathbb{R}}{
dt'
\int_{\mathbb{R}^3}{
\frac{q_s}{4 \pi \epsilon}
\frac{\vec{v}_s(t')}{\left|\vec{r}'-\vec{r}_s(t')\right|}
G(\vec{r},t;\vec{r}',t')
dV'
}
}
\nonumber
\end{align}

\noindent
In the physical setting shown in Fig. \ref{fig2} the coordinates of the test
charge $q$ are fixed, hence, order in which we apply operator 
$\nabla\times\nabla\times$ and second order time derivative 
$\frac{\partial}{\partial t^2}$ can be swapped (because operator $\nabla$ 
does not affect variable $t$). Furthermore, because variables $t'$, $x'$, $y'$
and $z'$ are independent of time $t$ we can move the double time derivative
under the integral sign in the last right hand side integral of above equation:

\begin{align}
\label{eq_coulomb_d12}
\frac{q_s}{4 \pi \epsilon}
\frac{\partial}{\partial t}
\frac{\vec{r}-\vec{r}_s(t)}
{\left|\vec{r}-\vec{r}_s(t)\right|^3}
=
&-\frac{\partial}{\partial t} \nabla \int_{\mathbb{R}}{
dt'
\int_{\mathbb{R}^3}{
\frac{q_s}{\epsilon}
\delta\left(\vec{r'}-\vec{r}_s(t')\right)
G(\vec{r},t;\vec{r}',t')
dV'
}
}
\\
&-
\frac{1}{c^2}
\frac{\partial^2}{\partial t^2}
\int_{\mathbb{R}}{
dt'
\int_{\mathbb{R}^3}{
\frac{q_s}{\epsilon} \vec{v}_s(t') \delta\left(\vec{r}'-\vec{r}_s(t')\right)
G(\vec{r},t;\vec{r}',t')
dV'
}
}
\nonumber
\\
&+
\nabla\times\nabla\times
\int_{\mathbb{R}}{
dt'
\int_{\mathbb{R}^3}{
\frac{q_s}{4 \pi \epsilon}
\frac{\vec{v}_s(t')}{\left|\vec{r}'-\vec{r}_s(t')\right|}
\frac{1}{c^2}
\frac{\partial^2}{\partial t^2}
G(\vec{r},t;\vec{r}',t')
dV'
}
}
\nonumber
\end{align}

\noindent
The second order time derivative of $G(\vec{r},t;\vec{r}',t')$ in the last
term of equation (\ref{eq_coulomb_d12}) can be replaced with equation
(\ref{eq_inhom_wave}) to obtain:

\begin{align}
\label{eq_coulomb_d14}
\frac{q_s}{4 \pi \epsilon}
\frac{\partial}{\partial t}
\frac{\vec{r}-\vec{r}_s(t)}
{\left|\vec{r}-\vec{r}_s(t)\right|^3}
=
&-\frac{\partial}{\partial t} \nabla \int_{\mathbb{R}}{
dt'
\int_{\mathbb{R}^3}{
\frac{q_s}{\epsilon}
\delta\left(\vec{r'}-\vec{r}_s(t')\right)
G(\vec{r},t;\vec{r}',t')
dV'
}
}
\\
&-
\frac{1}{c^2}
\frac{\partial^2}{\partial t^2}
\int_{\mathbb{R}}{
dt'
\int_{\mathbb{R}^3}{
\frac{q_s}{\epsilon} \vec{v}_s(t') \delta\left(\vec{r}'-\vec{r}_s(t')\right)
G(\vec{r},t;\vec{r}',t')
dV'
}
}
\nonumber
\\
&+
\nabla\times\nabla\times
\int_{\mathbb{R}}{
dt'
\int_{\mathbb{R}^3}{
\frac{q_s}{4 \pi \epsilon}
\frac{\vec{v}_s(t')}{\left|\vec{r}'-\vec{r}_s(t')\right|}
\delta\left(\vec{r}-\vec{r}'\right)\delta\left(t-t'\right)
dV'
}
}
\nonumber
\\
&+
\nabla\times\nabla\times
\int_{\mathbb{R}}{
dt'
\int_{\mathbb{R}^3}{
\frac{q_s}{4 \pi \epsilon}
\frac{\vec{v}_s(t')}{\left|\vec{r}'-\vec{r}_s(t')\right|}
\nabla^2
G(\vec{r},t;\vec{r}',t')
dV'
}
}
\nonumber
\end{align}

\noindent
Using sifting property of Dirac's delta function allows us to rewrite the
third right hand side term of equation (\ref{eq_coulomb_d14}) as:

\begin{eqnarray}
\label{eq_coulomb_d15}
\lefteqn{
\nabla\times\nabla\times
\int_{\mathbb{R}}{
dt'
\int_{\mathbb{R}^3}{
\frac{q_s}{4 \pi \epsilon}
\frac{\vec{v}_s(t')}{\left|\vec{r}'-\vec{r}_s(t')\right|}
\delta\left(\vec{r}-\vec{r}'\right)\delta\left(t-t'\right)
dV'
}
}
=
}
\\
&=&
\nabla\times\nabla\times
\int_{\mathbb{R}}{
dt'
\frac{q_s}{4 \pi \epsilon}
\frac{\vec{v}_s(t')}{\left|\vec{r}-\vec{r}_s(t')\right|}
\delta\left(t-t'\right)
}
=
\nonumber
\\
&=&
\nabla\times\nabla\times 
\frac{q_s}{4 \pi \epsilon}
\frac{\vec{v}_s(t)}{\left|\vec{r}-\vec{r}_s(t)\right|}
\nonumber
\end{eqnarray}

\noindent
To continue the derivation of Maxwell's equations from Coulomb's law we should
note that operator $\nabla$ does not affect vector $\vec{v}_s(t)$ because 
$\vec{v}_s(t)$ is a function of variable $t$. Hence, the application of
standard vector calculus identity 
$\nabla\times\nabla\times\vec{P}=\nabla\left(\nabla\cdot\vec{P}\right)-\nabla^2\vec{P}$  
yields:

\begin{align}
\label{eq_coulomb_d16}
\nabla\times\nabla\times 
\frac{q_s}{4 \pi \epsilon}
\frac{\vec{v}_s(t)}{\left|\vec{r}-\vec{r}_s(t)\right|}
&=
\frac{q_s}{4 \pi \epsilon}
\nabla
\left(
\nabla\cdot
\frac{\vec{v}_s(t)}{\left|\vec{r}-\vec{r}_s(t)\right|}
\right)
-
\frac{q_s}{4 \pi \epsilon}
\nabla^2
\frac{\vec{v}_s(t)}{\left|\vec{r}-\vec{r}_s(t)\right|}
=
\\
&=
\frac{q_s}{4 \pi \epsilon}
\nabla
\left(
\vec{v}_s(t)
\cdot
\nabla
\frac{1}{\left|\vec{r}-\vec{r}_s(t)\right|}
\right)
-
\frac{q_s}{4 \pi \epsilon}
\vec{v}_s(t)
\nabla^2
\frac{1}{\left|\vec{r}-\vec{r}_s(t)\right|}
=
\nonumber
\\
&=
-
\frac{q_s}{4 \pi \epsilon}
\nabla
\frac{\partial}{\partial t}
\frac{1}{\left|\vec{r}-\vec{r}_s(t)\right|}
+
\frac{q_s}{\epsilon}\vec{v}_s(t)\delta\left(\vec{r}-\vec{r}_s(t)\right)
=
\nonumber
\\
&=
\frac{q_s}{4 \pi \epsilon}
\frac{\partial}{\partial t}
\frac{\vec{r}-\vec{r}_s(t)}{\left|\vec{r}-\vec{r}_s(t)\right|^3}
+
\frac{q_s}{\epsilon}\vec{v}_s(t)\delta\left(\vec{r}-\vec{r}_s(t)\right)
\nonumber
\end{align}

\noindent
Combining equations (\ref{eq_coulomb_d14}), (\ref{eq_coulomb_d15}) and
(\ref{eq_coulomb_d16}), after cancellation of appropriate terms, yields:

\begin{align}
\label{eq_coulomb_d17}
0 = &- 
\frac{\partial}{\partial t} \nabla \int_{\mathbb{R}}{
dt'
\int_{\mathbb{R}^3}{
\frac{q_s}{\epsilon}
\delta\left(\vec{r'}-\vec{r}_s(t')\right)
G(\vec{r},t;\vec{r}',t')
dV'
}
}
\\
&-
\frac{1}{c^2}
\frac{\partial^2}{\partial t^2}
\int_{\mathbb{R}}{
dt'
\int_{\mathbb{R}^3}{
\frac{q_s}{\epsilon} \vec{v}_s(t') \delta\left(\vec{r}'-\vec{r}_s(t')\right)
G(\vec{r},t;\vec{r}',t')
dV'
}
}
\nonumber
\\
&+
\frac{q_s}{\epsilon}\vec{v}_s(t)\delta\left(\vec{r}-\vec{r}_s(t)\right)
\nonumber
\\
&+
\nabla\times\nabla\times
\int_{\mathbb{R}}{
dt'
\int_{\mathbb{R}^3}{
\frac{q_s}{4 \pi \epsilon}
\frac{\vec{v}_s(t')}{\left|\vec{r}'-\vec{r}_s(t')\right|}
\nabla^2
G(\vec{r},t;\vec{r}',t')
dV'
}
}
\nonumber
\end{align}

\noindent
In Appendix \ref{sec:apx_identities}, subsection \ref{sec:apx_identities_d3},
we have derived the following mathematical identity:

\begin{eqnarray}
\label{eq_coulomb_d18}
\lefteqn{
\int_{\mathbb{R}}{
dt'
\int_{\mathbb{R}^3}{
\frac{q_s}{4 \pi \epsilon}
\frac{\vec{v}_s(t')}{\left|\vec{r}'-\vec{r}_s(t')\right|}
\nabla^2
G(\vec{r},t;\vec{r}',t')
dV'
}
}
=
}
\\
&=&
-
\int_{\mathbb{R}}{
dt'
\int_{\mathbb{R}^3}
{
\frac{q_s}{\epsilon}
\vec{v}_s(t')
\delta\left(\vec{r}'-\vec{r}_s(t')\right)
G(\vec{r},t;\vec{r}',t')
dV'
}
}
\nonumber
\end{eqnarray}

\noindent
By inserting equation (\ref{eq_coulomb_d18}) into equation
(\ref{eq_coulomb_d17}) it is obtained that:

\begin{align}
\label{eq_coulomb_d19}
0 = &- 
\frac{\partial}{\partial t} \nabla \int_{\mathbb{R}}{
dt'
\int_{\mathbb{R}^3}{
\frac{q_s}{\epsilon}
\delta\left(\vec{r'}-\vec{r}_s(t')\right)
G(\vec{r},t;\vec{r}',t')
dV'
}
}
\\
&-
\frac{1}{c^2}
\frac{\partial^2}{\partial t^2}
\int_{\mathbb{R}}{
dt'
\int_{\mathbb{R}^3}{
\frac{q_s}{\epsilon} \vec{v}_s(t') \delta\left(\vec{r}'-\vec{r}_s(t')\right)
G(\vec{r},t;\vec{r}',t')
dV'
}
}
\nonumber
\\
&+
\frac{q_s}{\epsilon}\vec{v}_s(t)\delta\left(\vec{r}-\vec{r}_s(t)\right)
\nonumber
\\
&-
\nabla\times\nabla\times
\int_{\mathbb{R}}{
dt'
\int_{\mathbb{R}^3}
{
\frac{q_s}{\epsilon}
\vec{v}_s(t')
\delta\left(\vec{r}'-\vec{r}_s(t')\right)
G(\vec{r},t;\vec{r}',t')
dV'
}
}
\nonumber
\end{align}

\noindent
If we now introduce new constant $\mu=\frac{1}{c^2\epsilon}$  and 
divide whole equation (\ref{eq_coulomb_d19}) by $c^2$ we obtain:

\begin{align}
\label{eq_coulomb_d19_2}
0 = &- \frac{1}{c^2}
\frac{\partial}{\partial t} \nabla \int_{\mathbb{R}}{
dt'
\int_{\mathbb{R}^3}{
\frac{q_s}{\epsilon}
\delta\left(\vec{r'}-\vec{r}_s(t')\right)
G(\vec{r},t;\vec{r}',t')
dV'
}
}
\\
&-
\frac{1}{c^2}
\frac{\partial^2}{\partial t^2}
\int_{\mathbb{R}}{
dt'
\int_{\mathbb{R}^3}{
q_s \mu c \frac{\vec{v}_s(t')}{c} \delta\left(\vec{r}'-\vec{r}_s(t')\right)
G(\vec{r},t;\vec{r}',t')
dV'
}
}
\nonumber
\\
&+
q_s\mu\vec{v}_s(t)\delta\left(\vec{r}-\vec{r}_s(t)\right)
\nonumber
\\
&-
\nabla\times\nabla\times
\int_{\mathbb{R}}{
dt'
\int_{\mathbb{R}^3}
{
q_s\mu c
\frac{\vec{v}_s(t')}{c}
\delta\left(\vec{r}'-\vec{r}_s(t')\right)
G(\vec{r},t;\vec{r}',t')
dV'
}
}
\nonumber
\end{align}

\noindent
Although it is perhaps not yet apparent, equation (\ref{eq_coulomb_d19}) is
Maxwell-Ampere equation given in introductory part of this paper as equation
(\ref{eq_maxwell_4}). To evaluate right hand side integrals in equation
(\ref{eq_coulomb_d19_2}) we use sifting property of Dirac's delta function 
$\int_{\mathbb{R}^3}{\delta\left(\vec{r}'-\vec{r}_s(t')\right)f(\vec{r}')dV'}=f(\vec{r}_s(t'))$, 
where $f(\vec{r}')$ is function of position vector $\vec{r}'$, to
rewrite the right hand side integrals in equation (\ref{eq_coulomb_d19_2}) as:

\begin{align}
\label{eq_coulomb_d20}
\int_{\mathbb{R}}{
dt'
\int_{\mathbb{R}^3}{
\frac{q_s}{\epsilon}  \delta\left(\vec{r}'-\vec{r}_s(t')\right)
G(\vec{r},t;\vec{r}',t')
dV'
}
}
&=
\int_{\mathbb{R}}
{
\frac{q_s}{\epsilon}
G(\vec{r},t;\vec{r}_s(t'),t')
dt'
}
\\
\label{eq_coulomb_d21}
\int_{\mathbb{R}}{
dt'
\int_{\mathbb{R}^3}{
q_s\mu c \frac{\vec{v}_s(t')}{c} \delta\left(\vec{r}'-\vec{r}_s(t')\right)
G(\vec{r},t;\vec{r}',t')
dV'
}
}
&=
\int_{\mathbb{R}}
{
q_s \mu c
\frac{\vec{v}_s(t')}{c}
G(\vec{r},t;\vec{r}_s(t'),t')
dt'
}
\end{align}

\noindent
To evaluate right hand side integrals in equations above we now replace
 Green's function $G(\vec{r},t;\vec{r}_s(t'),t')$ in these equations
 with equation
(\ref{eq_fund_wave}) to obtain:

\begin{align}
\label{eq_coulomb_d22}
\int_{\mathbb{R}}{
dt'
\int_{\mathbb{R}^3}{
\frac{q_s}{\epsilon}  \delta\left(\vec{r}'-\vec{r}_s(t')\right)
G(\vec{r},t;\vec{r}',t')
dV'
}
}
&=
\\
&\hspace{-5em}=
\int_{\mathbb{R}}
{
\frac{q_s}{\epsilon}
\frac{
\delta\left( t'-t+\frac{\left|\vec{r}-\vec{r}_s(t')\right|}{c}\right)
}{
4\pi \left|\vec{r}-\vec{r}_s(t')\right|
}
dt'
}
\nonumber
\\
\label{eq_coulomb_d23}
\int_{\mathbb{R}}{
dt'
\int_{\mathbb{R}^3}{
q_s\mu c \frac{\vec{v}_s(t')}{c} \delta\left(\vec{r}'-\vec{r}_s(t')\right)
G(\vec{r},t;\vec{r}',t')
dV'
}
}
&=
\\
&\hspace{-5em}=
\int_{\mathbb{R}}
{
q_s\mu c 
\frac{\vec{v}_s(t')}{c}
\frac{
\delta\left( t'-t+\frac{\left|\vec{r}-\vec{r}_s(t')\right|}{c}\right)
}{
4\pi \left|\vec{r}-\vec{r}_s(t')\right|
}
dt'
}
\nonumber
\end{align}

\noindent
The right hand side integrals in equations (\ref{eq_coulomb_d22}) and (\ref{eq_coulomb_d23}) can be evaluated by making use of the
following standard mathematical identity involving Dirac's delta function:

\begin{equation}
\label{eq_coulomb_d24}
\delta\left(f(u)\right)=\frac{\delta(u-u_0)}{\left|\frac{\partial}{\partial u}f(u)\right|_{u=u_0}}
\end{equation}

\noindent
where $f(u)$ is real function of real argument $u$, and $u_0$ is the solution
of equation $f(u_0)=0$. Using identity (\ref{eq_coulomb_d24}),  the Dirac's
delta function in equations (\ref{eq_coulomb_d22}) and (\ref{eq_coulomb_d23})
can be written as as:

\begin{align}
\label{eq_coulomb_d25}
\delta\left( t'-t+\frac{\left|\vec{r}-\vec{r}_s(t')\right|}{c}\right)
&=
\frac{
\delta(t'-t_r)
}{
\left|
1
-
\frac{1}{c}
\frac{\vec{v_s}(t')\cdot\left(\vec{r}-\vec{r}_s(t')\right)}
{
\left|\vec{r}-\vec{r}_s(t')\right|
}
\right|
}
=
\frac{\delta(t'-t_r)}
{
1-\frac{1}{c}
\frac{\vec{v_s}(t')\cdot\left(\vec{r}-\vec{r}_s(t')\right)}
{
\left|\vec{r}-\vec{r}_s(t')\right|
}
}
\\
&=
\frac{\delta(t'-t_r)}
{
1-\vec{\beta}(t_r)\cdot\vec{n}(t_r)
}
\nonumber
\end{align} 

\noindent
where $\vec{\beta}(t_r)$ and $\vec{n}(t_r)$ are given by equations
(\ref{eq_vec_beta}) and (\ref{eq_vec_n}), respectively. From equation
(\ref{eq_coulomb_d24}) it follows that
the time $t_r$ is the solution to the following equation:

\begin{equation}
\label{eq_coulomb_d26}
t-t_r-\frac{\left|\vec{r}-\vec{r}_s(t_r)\right|}{c}=0
\end{equation}

\noindent
Evidently, the time $t_r$ is the time when the
disturbance created by moving source charge at position in space 
$\vec{r}_s(t_r)$ was created. This disturbance moves through the space with finite  velocity
$c$ and reaches the position $\vec{r}$ of the test charge at time $t$. 
In the electromagnetic literature this  time $t_r$
 is commonly known  as retarded time. 
 
 To proceed with derivation of Maxwell equations, we now insert equation
 (\ref{eq_coulomb_d25}) into equations (\ref{eq_coulomb_d22}) and
 (\ref{eq_coulomb_d23}), and evaluate the integrals over $t'$ using the sifting property of Dirac's delta
 function to obtain:

\begin{align}
\label{eq_coulomb_d27}
\int_{\mathbb{R}}{
dt'
\int_{\mathbb{R}^3}{
\frac{q_s}{\epsilon}  \delta\left(\vec{r}'-\vec{r}_s(t')\right)
G(\vec{r},t;\vec{r}',t')
dV'
}
}
&=
\frac{1}{4\pi\epsilon}
\frac{q_s}
{
\left|\vec{r}-\vec{r}_s(t_r)\right|
\left(1-\vec{\beta}(t_r)\cdot\vec{n}(t_r)\right)
}
\\
\label{eq_coulomb_d28}
\int_{\mathbb{R}}{
dt'
\int_{\mathbb{R}^3}{
q_s\mu c  \frac{\vec{v}_s(t')}{c} \delta\left(\vec{r}'-\vec{r}_s(t')\right)
G(\vec{r},t;\vec{r}',t')
dV'
}
}
&=
\frac{\mu c}{4\pi}
\frac{q_s\vec{\beta}_s(t_r)}
{
\left|\vec{r}-\vec{r}_s(t_r)\right|
\left(1-\vec{\beta}(t_r)\cdot\vec{n}(t_r)\right)
}
\end{align}

\noindent
By inserting equations (\ref{eq_coulomb_d27}) and (\ref{eq_coulomb_d28}) 
into equation (\ref{eq_coulomb_d19_2}), and rearranging, it is obtained:

\begin{align}
\label{eq_coulomb_d29}
\nabla\times\nabla\times
\frac{\mu c}{4\pi}
\frac{q_s\vec{\beta}_s(t_r)}
{
\left|\vec{r}-\vec{r}_s(t_r)\right|
\left(1-\vec{\beta}(t_r)\cdot\vec{n}(t_r)\right)
}
=
&
q_s\mu\vec{v}_s(t)\delta\left(\vec{r}-\vec{r}_s(t)\right)
\\
&-
\frac{1}{c^2}
\frac{\partial}{\partial t}
\nabla
\frac{1}{4\pi\epsilon}
\frac{q_s}
{
\left|\vec{r}-\vec{r}_s(t_r)\right|
\left(1-\vec{\beta}(t_r)\cdot\vec{n}(t_r)\right)
}
\nonumber
\\
&
-
\frac{1}{c^2}
\frac{\partial^2}{\partial t^2}
\frac{\mu c}{4\pi}
\frac{q_s\vec{\beta}_s(t_r)}
{
\left|\vec{r}-\vec{r}_s(t_r)\right|
\left(1-\vec{\beta}(t_r)\cdot\vec{n}(t_r)\right)
}
\nonumber
\end{align}

\noindent
The first right hand side term of the equation above can be identified as 
the current $\vec{J}$ of the point charge distribution moving with velocity 
$\vec{v}_s(t)$ multiplied by constant $\mu$:

\begin{equation}
\label{eq_coulomb_d30}
\mu\vec{J}=\mu q_s\vec{v}_s(t)\delta\left(\vec{r}-\vec{r}_s(t)\right)
\end{equation}

\noindent
We now define scalar function $\theta(\vec{r},t)$ and vector function 
$\vec{Q}(\vec{r},t)$ as:

\begin{align}
\label{eq_coulomb_d31}
\theta(\vec{r},t) &= 
\frac{1}{4\pi\epsilon}
\frac{q_s}
{
\left|\vec{r}-\vec{r}_s(t_r)\right|
\left(1-\vec{\beta}(t_r)\cdot\vec{n}(t_r)\right)
}
\\
\label{eq_coulomb_d32}
\vec{Q}(\vec{r},t)
&=
\frac{\mu c}{4\pi}
\frac{q_s\vec{\beta}_s(t_r)}
{
\left|\vec{r}-\vec{r}_s(t_r)\right|
\left(1-\vec{\beta}(t_r)\cdot\vec{n}(t_r)\right)
}
\end{align}

\noindent
With the aid of scalar function $\theta(\vec{r},t)$, vector function 
$\vec{Q}(\vec{r},t)$, and expression $\mu\vec{J}$ given by equation
(\ref{eq_coulomb_d30}) the  equation (\ref{eq_coulomb_d29}) can be 
written as:

\begin{equation}
\label{eq_coulomb_d33}
\nabla\times\nabla\times\vec{Q}(\vec{r},t)
= \mu\vec{J} - \frac{1}{c^2}\frac{\partial}{\partial t}
\left(
-\nabla\theta(\vec{r},t)-\frac{\partial}{\partial t}\vec{Q}(\vec{r},t)
\right)
\end{equation}

\noindent
Furthermore, we now define two vector functions $\vec{M}$ and $\vec{N}$ as:

\begin{align}
\label{eq_coulomb_d34}
\vec{M}
&=
\nabla\times
\vec{Q}(\vec{r},t)
\\
\label{eq_coulomb_d35}
\vec{N}
&=
-
\nabla
\theta(\vec{r},t)
-
\frac{\partial}{\partial t}
\vec{Q}(\vec{r},t)
\end{align}

\noindent
Using definitions of vector functions $\vec{M}$ and $\vec{N}$ given by
equations (\ref{eq_coulomb_d34}) and (\ref{eq_coulomb_d35}) we can rewrite
equation (\ref{eq_coulomb_d33}) as:

\begin{equation}
\label{eq_coulomb_d36}
\nabla\times\vec{M} = \vec{J}+\frac{1}{c^2}\frac{\partial\vec{N}}{\partial t}
\end{equation}

\noindent
We shall now investigate the mathematical properties of vector fields $\vec{M}$ and $\vec{N}$. Note that
because for any differentiable vector field $\vec{P}$ we can write 
$\nabla\cdot\nabla\times\vec{P}=0$, from equation (\ref{eq_coulomb_d34}) 
it follows that:

\begin{equation}
\label{eq_coulomb_d37}
\nabla\cdot\vec{M} = 0
\end{equation}

\noindent
The curl of the gradient of any differentiable scalar function $\psi$  is
equal to zero, i.e. $\nabla\times\nabla\psi=0$.  Thus, taking the curl of 
equation (\ref{eq_coulomb_d35}) yields:

\begin{equation}
\label{eq_coulomb_d38}
\nabla\times\vec{N} = 
-\nabla\times\frac{\partial}{\partial t}\vec{Q}(\vec{r},t)=
-\frac{\partial}{\partial t}\nabla\times\vec{Q}(\vec{r},t)=
-\frac{\partial\vec{M}}{\partial t}
\end{equation}

\noindent
Finally, in Appendix \ref{sec:apx_identities}, subsection
\ref{sec:apx_identities_d4}, we have shown that the divergence of vector 
field $\vec{N}$ is:

\begin{equation}
\label{eq_coulomb_d39}
\nabla\cdot\vec{N}
=
\frac{q_s}{\epsilon}\delta\left(\vec{r}-\vec{r}_s(t)\right)
=
\frac{\rho(\vec{r},t)}{\epsilon}
\end{equation}

\noindent
which completes the derivation of electrodynamic equations from Coulomb's law.

To compare these equations to Maxwell's equations,
in Table \ref{table_coulomb_fields} we have summarized governing equations for
scalar potential $\theta(\vec{r},t)$, vector potential $\vec{Q}(\vec{r},t)$,
vector field $\vec{M}$ and vector field $\vec{N}$ which are all derived from
Coulomb's law. By comparison with Liénard–Wiechert potentials given in Table
\ref{table_maxwell_fields}, we see  that scalar potential $\theta(\vec{r},t)$
is identical to Liénard–Wiechert scalar potential $\phi(\vec{r},t)$ and vector
potential $\vec{Q}(\vec{r},t)$ is identical to Liénard–Wiechert magnetic vector
potential $\vec{A}(\vec{r},t)$. Furthermore, by comparing Table
\ref{table_coulomb_fields} and Table \ref{table_maxwell_fields} we find that
 vector field $\vec{M}$ is identical to magnetic flux density $\vec{B}$ and
 that vector field $\vec{N}$ is identical to electric field $\vec{E}$. 
 
 In Table \ref{table_diff_equations} we have compared Maxwell's equations
governing fields $\vec{B}$ and $\vec{E}$ with differential equations governing
vector fields $\vec{M}$ and $\vec{N}$.
Clearly, left hand side of Table \ref{table_diff_equations} is identical in
the mathematical form to the right hand side of the same table, hence,
differential equations governing vector fields $\vec{M}$ and $\vec{N}$ are
identical to those governing vector fields $\vec{B}$ and $\vec{E}$. This is
expected, because we already know that vector field $\vec{N}=\vec{E}$ and
vector field $\vec{M}=\vec{B}$.

Thus, it should be evident by now that we have derived Maxwell equations
and Liénard–Wiechert potentials
directly from Coulomb's law. This was achieved by mathematically relating known electrostatic
Coulomb's law acting on test charge at present time to  "unknown" electrodynamic
fields acting at past. The mathematical link between the
static case in the present and dynamic case in the past  was provided by
generalized Helmholtz theorem. The derived equations are valid for arbitrarily moving source charge and these equations are not confined to motions along straight line. Furthermore, it should be noted that we have derived the Maxwell equations and Liénard–Wiechert potentials directly from Coulomb's law without resorting to special relativity or Lorentz transformation.

\begin{table}[!h]
\caption{Potentials and vector fields derived from Coulomb's law.}
\label{table_coulomb_fields}
\vspace{.5em}
\centering
\begin{tabular}{p{0.1\textwidth}p{0.45\textwidth}p{0.35\textwidth}}
\rowcolor{gray!50}
{\bf \small symbol} & {\bf \small equation} &  {\bf \small description}
\xrowht[()]{10pt}
\\
\xrowht[()]{16pt}
$\theta(\vec{r},t)$ &
${\displaystyle \frac{1}{4\pi\epsilon}\frac{q_s}{\left|\vec{r}-\vec{r}_s(t_r)\right|\left(1-\vec{\beta}(t_r)\cdot\vec{n}(t_r)\right)}}$ & 
\vspace{-12pt}
\makecell{scalar potential derived from \\ Coulomb's law}
\\
\cdashline{1-3}[.4pt/1pt]
\\[-10pt]
\xrowht[()]{16pt}
$\vec{Q}(\vec{r},t)$
&
${\displaystyle \frac{\mu c}{4\pi}
\frac{q_s\vec{\beta}_s(t_r)}
{
\left|\vec{r}-\vec{r}_s(t_r)\right|
\left(1-\vec{\beta}(t_r)\cdot\vec{n}(t_r)\right)
}}$
&
\vspace{-12pt}
\makecell{vector potential derived from\\ Coulomb's law}
\\
\cdashline{1-3}[.4pt/1pt]
\\[-10pt]
\xrowht[()]{12pt}
$\vec{M}$
&
$\nabla\times\vec{Q}(\vec{r},t)$
&
\vspace{-12pt}
\makecell{vector field $\vec{M}$ derived from\\ Coulomb's law}
\\
\cdashline{1-3}[.4pt/1pt]
\\[-10pt]
\xrowht[()]{12pt}
$\vec{N}$
&
$-\nabla\phi(\vec{r},t)-\frac{\partial}{\partial t}\vec{Q}(\vec{r},t)$
&
\vspace{-12pt}
\makecell{vector field $\vec{N}$ derived from\\ Coulomb's law}
\\
\cdashline{1-3}[.4pt/1pt]
\end{tabular}
\end{table}

\begin{table}[!h]
\caption{Standard electromagnetic theory expressions for Liénard–Wiechert scalar potential $\phi(\vec{r},t)$,  Liénard–Wiechert vector potential $\vec{A}(\vec{r},t)$,  magnetic flux density $\vec{B}$ and electric field $\vec{E}$.}
\label{table_maxwell_fields}
\vspace{.5em}
\centering
\begin{tabular}{p{0.1\textwidth}p{0.45\textwidth}p{0.35\textwidth}}
\rowcolor{gray!50}
{\bf \small symbol} & {\bf \small equation} & {\bf \small description}
\xrowht[()]{10pt}
\\
\xrowht[()]{16pt}
$\phi(\vec{r},t)$ &
   ${\displaystyle \frac{1}{4\pi\epsilon}\frac{q_s}{\left|\vec{r}-\vec{r}_s(t_r)\right|\left(1-\vec{\beta}(t_r)\cdot\vec{n}(t_r)\right)}}$ 
  &  \makecell{Liénard–Wiechert scalar\\ potential}
\\
\cdashline{1-3}[.4pt/1pt]
\\[-10pt]
\xrowht[()]{16pt}
$\vec{A}(\vec{r},t)$
&
${\displaystyle \frac{\mu c}{4\pi}
\frac{q_s\vec{\beta}_s(t_r)}
{
\left|\vec{r}-\vec{r}_s(t_r)\right|
\left(1-\vec{\beta}(t_r)\cdot\vec{n}(t_r)\right)
}}$
&
\makecell{Liénard–Wiechert vector\\ potential}
\\
\cdashline{1-3}[.4pt/1pt]
\\[-10pt]
\xrowht[()]{12pt}
$\vec{B}$
&
$\nabla\times\vec{A}(\vec{r},t)$
&
magnetic flux density
\\
\cdashline{1-3}[.4pt/1pt]
\\[-10pt]
\xrowht[()]{12pt}
$\vec{E}$
&
$-\nabla\phi(\vec{r},t)-\frac{\partial}{\partial t}\vec{A}(\vec{r},t)$
&
electric field
\\
\cdashline{1-3}[.4pt/1pt]
\end{tabular}
\end{table}

\begin{table}[!h]
\caption{Maxwell equations for electromagnetic fields $\vec{E}$ and $\vec{B}$ compared with differential equations for vector fields $\vec{M}$ and $\vec{N}$ derived from Coulomb's law.}
\label{table_diff_equations}
\vspace{.5em}
\centering
\begin{tabular}{p{0.3\textwidth}p{0.3\textwidth}p{0.3\textwidth}}
\rowcolor{gray!50}
{\bf \small Maxwell equation} & {\bf \small description} & {\bf \small \makecell{equations derived from\\Coulomb's law}}
\xrowht[()]{10pt}
\\
\cdashline{1-3}[.4pt/1pt]
\\[-10pt]
\xrowht[()]{16pt}
$\nabla\times\vec{B}=\vec{J}+\frac{1}{c^2}\frac{\partial\vec{E}}{\partial t}$
&
Maxwell-Ampere equation
&
$\nabla\times\vec{M}=\vec{J}+\frac{1}{c^2}\frac{\partial\vec{N}}{\partial t}$
\\
\cdashline{1-3}[.4pt/1pt]
\\[-10pt]
\xrowht[()]{16pt}
$\nabla\times\vec{E}=-\frac{\partial\vec{B}}{\partial t}$
&
Faraday's law
&
$\nabla\times\vec{N}=-\frac{\partial\vec{M}}{\partial t}$
\\
\cdashline{1-3}[.4pt/1pt]
\\[-10pt]
\xrowht[()]{16pt}
$\nabla\cdot\vec{E}=\frac{\rho}{\epsilon}$
&
Gauss' law for electric field
&
$\nabla\cdot\vec{N}=\frac{\rho}{\epsilon}$
\\
\cdashline{1-3}[.4pt/1pt]
\\[-10pt]
\xrowht[()]{16pt}
$\nabla\cdot\vec{B}=0$
&
Gauss' law for magnetic field
&
$\nabla\cdot\vec{M}=0$
\\
\cdashline{1-3}[.4pt/1pt]
\end{tabular}
\end{table}

\section{Derivation of Electrodynamic Energy Conservation Law and Lorentz Force}\label{sec:lorentz_deriv}
\noindent
To derive the electrodynamic energy conservation law from Coulomb's law we
first  consider  hypothetical physical setting shown in Fig. \ref{fig3}
where the source charge $q_s$ is moving along arbitrary trajectory 
$\vec{r}_s(t)$. Then the  source charge $q_s$ stops at some time in the past $t_s$. In this physical setting, closed contour $C$  is at rest at
all times.   At
present time $t_p>t_s$ all the points inside the sphere of radius $R=c(t_p-t_s)$
 are affected only by electrostatic Coulomb's field. The known energy conservation law valid at present
dictates that  contour integral of electrostatic field along any closed
contour  immersed inside the sphere of radius $R$ equals to zero:

\begin{figure}[!t]
\centerline{\includegraphics[width=.8\textwidth]{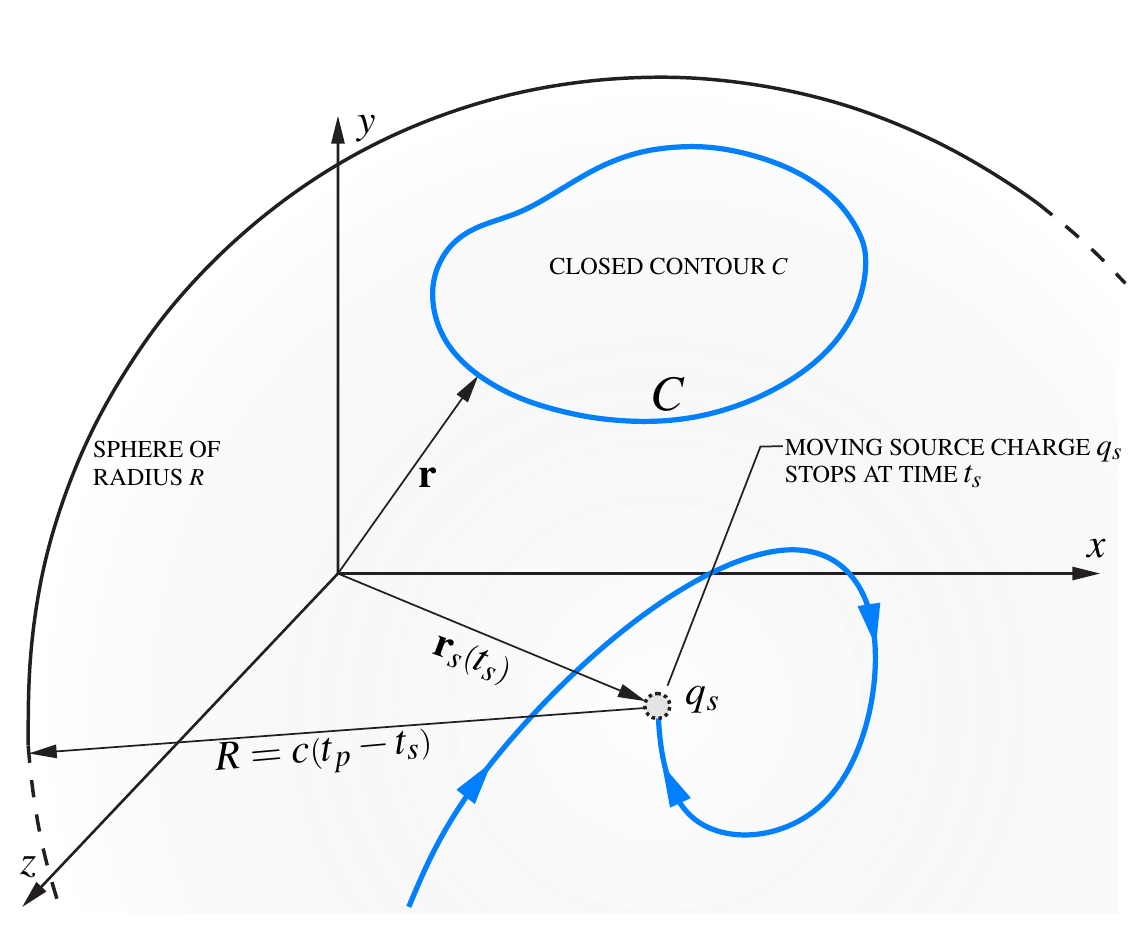}}
\caption{Source charge $q_s$ is moving along arbitrary trajectory 
$\vec{r}_s(t)$ and it stops at past time $t_s$. Closed contour $C$  is at rest
at all times. All the points $\vec{r}$ on contour $C$ are inside the sphere of
radius $R=c(t_p-t_s)$. At present time $t_p>t_s$ all the points on contour $C$
are affected only by Coulomb's electrostatic field.}
\label{fig3}
\end{figure}

\begin{equation}
\label{eq_lorentz_0}
\oint_{C}{\vec{E}_c(\vec{r},t_p)\cdot d\vec{r}}
=
\oint_{C}{\frac{q_s}{4\pi\epsilon}
\frac{\vec{r}-\vec{r}_s(t_s)}{\left|\vec{r}-\vec{r}_s(t_s)\right|^3}
\cdot
d\vec{r}
}
=0
\end{equation}

\noindent
where $\vec{E}_c$ is Coulomb's electrostatic field, $\vec{r}_s(t_s)$ is the
position vector of source charge when it  stopped moving, and vector $\vec{r}$
is  the position vector of the point on contour $C$. This electrostatic energy conservation law, valid at present time $t_p$,
states that no net work is done in transporting the unit charge  along any
closed contour immersed in electrostatic field.

To proceed, we assume that in the past, when the source charge was moving, the energy conservation law is unknown. However, generalized Helmholtz decomposition theorem allows us to derive this "unknown" electrodynamic energy conservation law valid in the past from the knowledge of electrostatic energy conservation law valid at present. To derive this unknown electrodynamic electrodynamic conservation law we
consider the contour integral (\ref{eq_lorentz_0}) at the moment $t$
infinitesimally before the  time when the source charge stopped:

\begin{equation}
\label{eq_lorentz_0_2}
t = t_s - \delta t
\end{equation}

\noindent
where $\delta t$ is infinitesimally small time interval.
If time interval $\delta t$ approaches zero ($\delta t\rightarrow 0$) we can
rewrite the contour integral (\ref{eq_lorentz_0}) as the function of time $t$:

\begin{equation}
\label{eq_lorentz_0_3}
\oint_{C}{\vec{E}_c(\vec{r},t_p-\delta t)\cdot d\vec{r}}
=
\oint_{C}{\frac{q_s}{4\pi\epsilon}
\frac{\vec{r}-\vec{r}_s(t)}{\left|\vec{r}-\vec{r}_s(t)\right|^3}
\cdot
d\vec{r}
}
=0
\end{equation}

\noindent
Because the integrand on the right hand side of equation
(\ref{eq_lorentz_0_3}) is the function of varying time $t$ and position vector
$\vec{r}$  the generalized Helmholtz decomposition theorem  can be applied to
rewrite this integrand as the function of past positions and velocities of the
source charge.  In fact, such  expression is already derived in previous
section as equation (\ref{eq_coulomb_d12}), repeated here for clarity:
 
 \begin{align}
\label{eq_lorentz_3}
\frac{q_s}{4 \pi \epsilon}
\frac{\vec{r}-\vec{r}_s(t)}
{\left|\vec{r}-\vec{r}_s(t)\right|^3}
=
&-\nabla \int_{\mathbb{R}}{
dt'
\int_{\mathbb{R}^3}{
\frac{q_s}{\epsilon}
\delta\left(\vec{r'}-\vec{r}_s(t')\right)
G(\vec{r},t;\vec{r}',t')
dV'
}
}
\\
&-
\frac{1}{c^2}
\frac{\partial}{\partial t}
\int_{\mathbb{R}}{
dt'
\int_{\mathbb{R}^3}{
\frac{q_s}{\epsilon} \vec{v}_s(t') \delta\left(\vec{r}'-\vec{r}_s(t')\right)
G(\vec{r},t;\vec{r}',t')
dV'
}
}
\nonumber
\\
&+
\frac{1}{c^2}
\frac{\partial}{\partial t}
\int_{\mathbb{R}}{
dt'
\int_{\mathbb{R}^3}{
\frac{q_s}{4 \pi \epsilon}
\left[
\nabla'
\times
\nabla'
\times
\frac{\vec{v}_s(t')}{\left|\vec{r}'-\vec{r}_s(t')\right|}
\right]
G(\vec{r},t;\vec{r}',t')
dV'
}
}
\nonumber
\end{align}

\noindent
Substituting the first two right hand side terms of equation
(\ref{eq_lorentz_3}) with equations (\ref{eq_coulomb_d27}) and
(\ref{eq_coulomb_d28}) and combining the result with equations
(\ref{eq_coulomb_d31}) and (\ref{eq_coulomb_d32}), and using 
$c^2=1/\mu\epsilon$ yields:

\begin{equation}
\label{eq_lorentz_4}
\frac{q_s}{4 \pi \epsilon}
\frac{\vec{r}-\vec{r}_s(t)}
{\left|\vec{r}-\vec{r}_s(t)\right|^3}
=
-\nabla\theta(\vec{r},t)
-\frac{\partial}{\partial t}
\vec{Q}(\vec{r},t)
+
\vec{K}(\vec{r},t)
\end{equation}

\noindent
where vector function $\vec{K}(\vec{r},t)$ is equal to the last right hand
side term of equation (\ref{eq_lorentz_3}):

\begin{equation}
\label{eq_lorentz_5}
\vec{K}(\vec{r},t)
=
\frac{1}{c^2}
\frac{\partial}{\partial t}
\int_{\mathbb{R}}{
dt'
\int_{\mathbb{R}^3}{
\frac{q_s}{4 \pi \epsilon}
\left[
\nabla'
\times
\nabla'
\times
\frac{\vec{v}_s(t')}{\left|\vec{r}'-\vec{r}_s(t')\right|}
\right]
G(\vec{r},t;\vec{r}',t')
dV'
}
}
\end{equation}

\noindent
Replacing the first two terms on the right hand side of equation
(\ref{eq_lorentz_4}) with equation (\ref{eq_coulomb_d35}) yields:

\begin{equation}
\label{eq_lorentz_7}
\frac{ q_s}{4 \pi \epsilon}
\frac{\vec{r}-\vec{r}_s(t)}
{\left|\vec{r}-\vec{r}_s(t)\right|^3}
=
\vec{N}(\vec{r},t)
+
\vec{K}(\vec{r},t)
\end{equation}

\noindent
Then, by inserting equation (\ref{eq_lorentz_7}) into right hand side of
equation (\ref{eq_lorentz_0_3}) it is obtained that:

\begin{equation}
\label{eq_lorentz_7_1}
0
=
\oint_{C}{\vec{E}_c(\vec{r},t_p-\delta t)\cdot d\vec{r}}
=
\oint_{C}{
\left(
\vec{N}(\vec{r},t)
+
\vec{K}(\vec{r},t)
\right)
\cdot
d\vec{r}
}
\end{equation}

\noindent
The space-time integral on the right hand side of equation
(\ref{eq_lorentz_5}) is very difficult to evaluate. However, we can eliminate
vector field $\vec{K}(\vec{r},t)$ from the right hand side of equation
(\ref{eq_lorentz_7_1}) by the application of Stokes' theorem:

\begin{equation}
\label{eq_lorentz_7_15}
0
=
\oint_{C}{\vec{E}_c(\vec{r},t_p-\delta t)\cdot d\vec{r}}
=
\oint_{C}{
\vec{N}(\vec{r},t)
\cdot
d\vec{r}}
+
\int_{S}{
\nabla\times\vec{K}(\vec{r},t)
\cdot
d\vec{S}
}
\end{equation}

\noindent
From here, we take the curl of both sides of equation (\ref{eq_lorentz_7}) and
by combining with equation (\ref{eq_coulomb_d38}) it is obtained that:

\begin{equation}
\label{eq_lorentz_8}
\nabla\times\vec{K}(\vec{r},t)
=
-\nabla\times\vec{N}(\vec{r},t)
=
\frac{\partial}{\partial t}\vec{M}(\vec{r},t)
\end{equation}

\noindent
Because surface $S$ and contour $C$ are stationary we can write that 
$\frac{\partial}{\partial t}\vec{M}(\vec{r},t)=\frac{d}{dt}\vec{M}(\vec{r},t)$.
Inserting equation (\ref{eq_lorentz_8}) into equation (\ref{eq_lorentz_7_15}) 
and taking into account that surface $S$ and contour $C$ are not moving
yields:

\begin{equation}
\label{eq_lorentz_8_1}
0
=
\oint_{C}{\vec{E}_c(\vec{r},t_p-\delta t)\cdot d\vec{r}}
=
\oint_{C}{
\vec{N}(\vec{r},t)
\cdot
d\vec{r}}
+
\frac{d}{d t}
\int_{S}{
\vec{M}(\vec{r},t)
\cdot
d\vec{S}
}
\end{equation}

\noindent 
The right hand side of equation (\ref{eq_lorentz_8_1}) is unknown energy
conservation principle valid for varying in time dynamic fields 
$\vec{N}(\vec{r},t)$ and $\vec{M}(\vec{r},t)$ and it is derived  from
electrostatic energy conservation principle valid at present time. If 
$\vec{N}$ is replaced by $\vec{E}$ and if $\vec{M}$ is replaced by $\vec{B}$
it can be seen that  we have just obtained the physical law known in
electrodynamics as Faraday's law. 

From equation (\ref{eq_lorentz_8_1}) the conclusion can be drawn about the
nature of Faraday's law. It represents the energy conservation principle valid
for non-conservative dynamic fields and it is dynamic equivalent of
electrostatic energy conservation principle valid for Coulomb's electrostatic  field.

\begin{figure}[!b]
\centerline{\includegraphics[width=.9\textwidth]{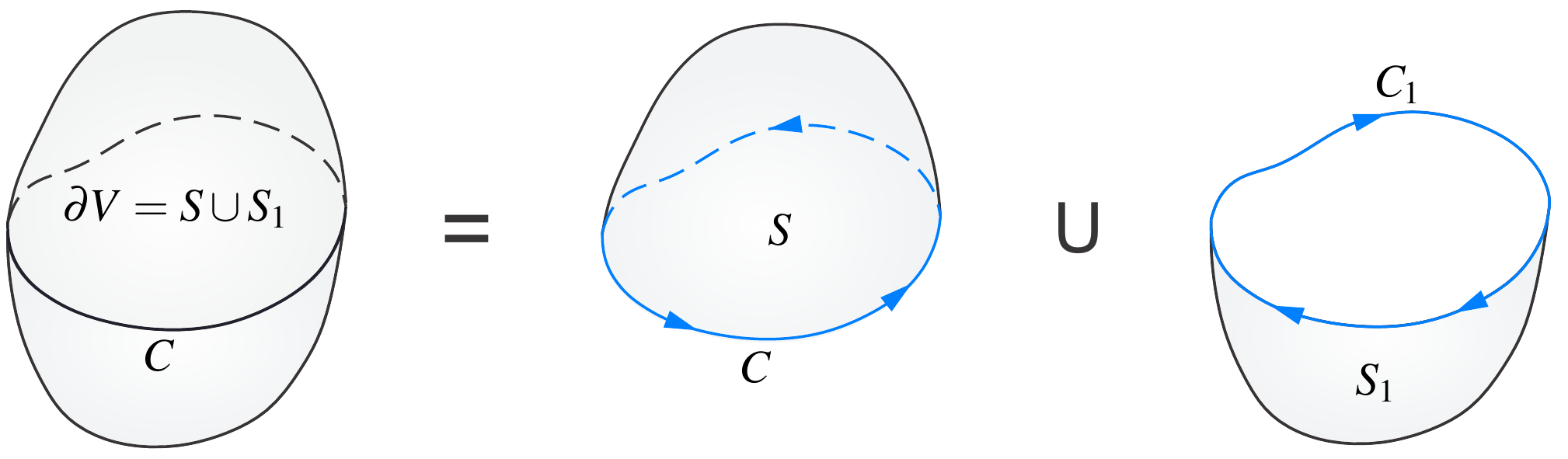}}
\caption{Closed surface $\partial V$ that bounds volume $V$ is union of two
surfaces $S$ and $S_1$. Contour $C$ bounds surface $S$ and contour $C_1$
bounds surface $S_1$. Contours $C$ and $C_1$ are identical, however they have
different Stokes' orientation.}
\label{fig4}
\end{figure}

However, even the Faraday's law itself can be considered as consequence of
something else.
To see this, consider simply connected volume $V$ bounded by surface 
$\partial V$ as shown in Fig. \ref{fig4}. The surface $\partial V$ is union
of two surfaces $S$ and $S_1$ bounded by respective contours $C$ and $C_1$.
Contours $C$ and $C_1$ consist of exactly the same spatial points, however,
the Stokes' orientation of these contours is opposite $C=-C_1$. Then, using 
$\nabla\times\vec{N}(\vec{r},t)=-\frac{\partial}{\partial t}\vec{M}(\vec{r},t)$ 
the first right hand side contour integral of equation (\ref{eq_lorentz_8_1})
can be written as:

\begin{equation}
\label{eq_lorentz_9}
\oint_{C}{
\vec{N}(\vec{r},t)
\cdot
d\vec{r}}
=
-
\oint_{C_1}{
\vec{N}(\vec{r},t)
\cdot
d\vec{r}}
=
-
\int_{S_1}{\nabla\times\vec{N}(\vec{r},t)\cdot d\vec{S}}
=
\frac{d}{dt}
\int_{S_1}{\vec{M}(\vec{r},t)\cdot d\vec{S}}
\end{equation}

\noindent
Replacing the first right hand side term of equation (\ref{eq_lorentz_8_1})
with equation (\ref{eq_lorentz_9}) yields different form of dynamic energy
conservation law:

\begin{equation}
\label{eq_lorentz_9_1}
0
=
\oint_{C}{\vec{E}_c(\vec{r},t_p-\delta t)\cdot d\vec{r}}
=
\frac{d}{d t}
\oint_{\partial V}{
\vec{M}(\vec{r},t)
\cdot
d\vec{S}
}
\end{equation}

\noindent
If we replace $\vec{M}(\vec{r},t)$ with $\vec{B}$ we see that right hand side
of equation (\ref{eq_lorentz_9_1}) is time derivative of Gauss' law for
magnetic fields. The standard interpretation of Gauss' law for magnetic fields
is that magnetic monopoles do not exist. However, from equation
(\ref{eq_lorentz_9_1}) we conclude that alternative interpretation of this law
is that its time derivative represents the dynamic energy conservation law.
From the derivations presented, we might even say that Faraday's law is the
consequence of Gauss' law for magnetic fields. It should be noted that these
energy-conservation equations were all derived from simple electrostatic
Coulomb's law.

From dynamic energy conservation law the derivation of Lorentz force is
straightforward: we now assume that all the points on surface $\partial V$
shown in Fig. \ref{fig4} have some definite velocity $\vec{v}$ such that
$|\vec{v}|<<c$. Then the surface $\partial V$ is the function of time, hence,
$C=C(t)$ and $S=S(t)$. Hence, we can rewrite equation (\ref{eq_lorentz_9_1}) as the sum of two surface integrals:

\begin{equation}
\label{eq_lorentz_9_1_1}
\frac{d}{d t}
\oint_{\partial V}{
\vec{M}(\vec{r},t)
\cdot
d\vec{S}
}
=
\frac{d}{d t}
\oint_{S(t)}{
\vec{M}(\vec{r},t)
\cdot
d\vec{S}
}
+
\frac{d}{d t}
\oint_{S_1(t)}{
\vec{M}(\vec{r},t)
\cdot
d\vec{S}
}
= 0
\end{equation}

\noindent
where $\partial V=S(t)\cup S_1(t)$.
The Leibniz identity \cite{Flanders1973}  for moving
surfaces states that for any differentiable vector field $\vec{P}$ we can
write:

\begin{equation}
\label{eq_lorentz_10}
\frac{d}{dt}
\int_{S(t)}{\vec{P}\cdot d\vec{S}}
=
\int_{S(t)}{\left[\frac{\partial}{\partial t}\vec{P} +
 \left(\nabla\cdot\vec{P}\right)\vec{v}\right]\cdot d\vec{S}}
-
\oint_{C(t)}{\vec{v}\times\vec{P}\cdot d\vec{r}}
\end{equation}

\noindent
Applying the Leibniz identity to the surface integral over surface $S_1(t)$
in equation (\ref{eq_lorentz_9_1_1}) and using $\nabla\cdot\vec{M}(\vec{r},t)=0$ yields:

\begin{equation}
\label{eq_lorentz_9_1_2}
\frac{d}{d t}
\oint_{S(t)}{
\vec{M}(\vec{r},t)
\cdot
d\vec{S}
}
+
\oint_{S_1(t)}{
\frac{\partial}{\partial t}
\vec{M}(\vec{r},t)
\cdot
d\vec{S}
}
-
\oint_{C_1(t)}{
\vec{v}\times\vec{M}(\vec{r},t)
\cdot d\vec{r}
}
= 0
\end{equation}

\noindent
Using the result from previous section, i.e. $\nabla\times\vec{N}(\vec{r},t)=-\frac{\partial}{\partial t}\vec{M}(\vec{r},t)$, and applying the Stokes' theorem yields:

\begin{equation}
\label{eq_lorentz_9_1_3}
\frac{d}{d t}
\oint_{S(t)}{
\vec{M}(\vec{r},t)
\cdot
d\vec{S}
}
-
\oint_{C_1(t)}{
\vec{N}(\vec{r},t)
\cdot
d\vec{r}
}
-
\oint_{C_1(t)}{
\vec{v}\times\vec{M}(\vec{r},t)
\cdot d\vec{r}
}
= 0
\end{equation}

\noindent
Because curves $C_1(t)$ and $C(t)$ comprise of same points, however,  Stokes' orientation of curves $C_1(t)$ and $C(t)$ is opposite, i.e. $C_1(t)=-C(t)$, we can rewrite equation (\ref{eq_lorentz_9_1_3}) as:

\begin{equation}
\label{eq_lorentz_11}
\oint_{C(t)}
{
\left[
\vec{N}(\vec{r},t)
+
\vec{v}\times\vec{M}(\vec{r},t)
\right]
\cdot d\vec{r}
}
+
\frac{d}{d t}
\int_{S(t)}{
\vec{M}(\vec{r},t)
\cdot
d\vec{S}
}
=
0
\end{equation}

\noindent
 Note that equation (\ref{eq_lorentz_11}) could not be derived from the right
 hand side of equation (\ref{eq_lorentz_8_1}), i.e. from Faraday's law, even with Leibniz rule. For that
 reason, we might take that the energy conservation law on the right hand side
 of equation (\ref{eq_lorentz_9_1}) is perhaps more general than the one given
 by equation (\ref{eq_lorentz_8_1}).

Furthermore, note that the time derivative of the surface integral in equation (\ref{eq_lorentz_11}) does not represent the work of any force.  However, from equation (\ref{eq_lorentz_9_1}) we know that the terms in equation (\ref{eq_lorentz_11}) have dimensions of the work done by electrodynamic force in moving the unit charge along contour $C(t)$. Because the first term in equation (\ref{eq_lorentz_11}) is contour integral of vector field we can conclude that this term represents the non-zero work  done by non-conservative electrodynamic force in transporting the unit charge along contour $C(t)$.

Hence,
just as the left hand side of equation (\ref{eq_lorentz_9_1}) represents the
work done by conservative electrostatic force in transporting the unit charge
along contour $C$, the contour integral on the left of equation
(\ref{eq_lorentz_11}) represents the work done by non-conservative
electrodynamic force in transporting the unit charge along the same contour.
The purpose of surface integral on the left hand side of equation
(\ref{eq_lorentz_11}) is to balance non-zero work of non-conservative
electrodynamic force along contour $C$. 
Thus, it can be concluded that the electrodynamic force $\vec{F}_D$ on  charge
$q$ moving with velocity $\vec{v}$ along contour $C$ is:

\begin{equation}
\vec{F}_D = q \vec{N}(\vec{r},t)
+
q \vec{v}\times\vec{M}(\vec{r},t)
\end{equation}

\noindent
Finally, in previous section we have shown that $\vec{N}=\vec{E}$ and that $\vec{M}=\vec{B}$. Thus, by replacing $\vec{N}$ with $\vec{E}$ and $\vec{M}$ with $\vec{B}$ it is obtained that:

\begin{equation}
\vec{F}_D = q \left(\vec{E}+\vec{v}\times\vec{B}\right)
\end{equation}

\noindent
which is expression for well known Lorentz force. 
It was derived theoretically from the knowledge of electrostatic energy conservation law which, in turn, can be derived from Coulomb's law. Thus, we may say that we have just derived the Lorentz force  from simple electrostatic Coulomb's law.

\section{Conclusion}
In this paper we have presented the theoretical framework that explains Maxwell equations and the Lorentz force on more fundamental level than it was  previously done. Maxwell derived Maxwell equations from experimental Ampere's force law and experimental Faraday's law, and Lorentz continued work on Maxwell's theory to discover the Lorentz force. In last 150 years,  no successful theory was presented that would explain Maxwell's equations and Lorentz force on more fundamental level.

To accomplish this, relativistically correct Liénard–Wiechert potentials, Maxwell
equations and the Lorentz force were derived directly from electrostatic
Coulomb's law. 
 In contrast to frequently criticized previous attempts to derive Maxwell's
 equations from Coulomb's law using special relativity and Lorentz
 transformation, the Lorentz transformation was not used in our derivations 
 nor the theory of special relativity. In fact, in this work, dynamic 
 Liénard–Wiechert potentials, Maxwell equations and Lorentz force were derived
 from Coulomb's law using the following two simple postulates:

\begin{enumerate}[label=(\alph*)]
\item when charges are at rest the Coulomb's law describes the force acting 
between charges
\vspace{.5em}
\item disturbances caused by moving charges propagate outwardly from moving
charge with finite velocity
\end{enumerate}

\noindent
The derivation of these dynamic physical laws from electrostatic Coulomb's law
would not be possible without generalized Helmholtz decomposition theorem also
derived in this paper. This theorem allows the vector function of present
position and present time to be written as space-time integral of positions
and velocities at previous time. In contrast, standard Helmholtz decomposition
theorem is valid for functions of space only and it ignores time.

To derive the Lorentz force from Coulomb's law, in section \ref{sec:lorentz_deriv}, the "unknown" dynamic energy conservation
law valid in the past was derived from the knowledge of electrostatic energy
conservation law valid at present. The link between the present and the past
was again provided by generalized Helmholtz decomposition theorem. This
"unknown" dynamic energy conservation principle turned out to be Faraday's law
of induction. Additionally, it was shown that Faraday's law of induction can
be considered equivalent to time derivative of Gauss' law for magnetic field. From these energy
conservation considerations the
Lorentz force was derived.

From the presented analysis  one  important question naturally arises: are
Maxwell's equations and Lorentz force the consequence of electrostatic
Coulomb's law? They are most probably not. It is rather the opposite,
Coulomb's law is the limiting case of  Lorentz force when the source charge
becomes stationary. However, as it was shown in this paper, it is entirely
possible to deduce dynamic Maxwell equations and Lorentz force from the
knowledge of simple electrostatic Coulomb's law. 

Finally, this paper attempts to answer another important question:
how can we deduce more general dynamic physical laws from
the limited knowledge provided by static physical law? 
The significance of answering this question is that in the future it will
perhaps become possible  that similar reasoning could deepen the understanding
of physical laws other than Maxwell equations and Lorentz force.

\appendix
\renewcommand{\theequation}{\Alph{section}.\arabic{subsection}.\arabic{equation}}
\renewcommand{\thesubsection}{\Alph{section}.\arabic{subsection}}
\section{Derivation of generalized Helmholtz decomposition theorem}\label{sec:apx_gen_helmholtz}
\noindent
In this appendix, we derive the generalized Helmholtz decomposition theorem
for vector functions of space and time. However, in effort to enhance the
readability of this work, we first start by considering some basic identities
given in section \ref{sec:apx_identities_prelim} of this appendix. 

\subsection{Preliminary considerations}\label{sec:apx_identities_prelim}
\noindent
To clarify notation used throughout this paper we first define position vectors $\vec{r}$ and $\vec{r}'$ as:

\begin{align}
\vec{r} &= x \vec{\hat{x}} + y \vec{\hat{y}} + z \vec{\hat{z}} 
\\
\vec{r}' &= x' \vec{\hat{x}} + y' \vec{\hat{y}} + z' \vec{\hat{z}} 
\end{align}

\noindent
where $\vec{\hat{x}}$, $\vec{\hat{y}}$ and $\vec{\hat{z}}$ are Cartesian, mutually orthogonal, unit basis vectors. Variables $x,y,z\in\mathbb{R}$ and $x',y',z'\in\mathbb{R}$ are linearly independent variables. Furthermore, throughout this paper we use position vector $\vec{r}_s(t')$ to indicate the position of the source charge. This position vector $\vec{r}_s(t')$ is defined as:

\begin{equation}
\vec{r}_s(t') =  x_s(t') \vec{\hat{x}} + y_s(t') \vec{\hat{y}} + z_s(t') \vec{\hat{z}} 
\end{equation}

\noindent
where $x_s(t')$, $y_s(t')$ and $z_s(t')$ are all functions of real variable $t'\in\mathbb{R}$ which is independent of variables $x,y,z\in\mathbb{R}$ and $x',y',z'\in\mathbb{R}$. The time derivative of position vector $\vec{r}_s(t')$ is velocity $\vec{v}_s(t')$ of the source charge:

\begin{equation}
\label{eq_prelim_4}
\vec{v}_s(t') = \frac{\partial\vec{r}_s(t')}{\partial t'}
\end{equation}

\noindent
On many occasions in this paper we have used differential operators $\nabla$ and $\nabla'$ defined as:

\begin{align}
\nabla &= \vec{\hat{x}}\frac{\partial}{\partial x} +  \vec{\hat{y}}\frac{\partial}{\partial x} +  \vec{\hat{z}}\frac{\partial}{\partial x} 
\\
\nabla' &=  \vec{\hat{x}}\frac{\partial}{\partial x'} + \vec{\hat{y}}\frac{\partial}{\partial y'} + \vec{\hat{z}}\frac{\partial}{\partial z'} 
\end{align}

\noindent
Operator $\nabla$ acts only on functions of variables $x,y,z$, hence, on functions of position vector $\vec{r}$. On the other hand, operator $\nabla'$ acts only on functions of variables $x',y',z'$, thus, it acts on functions of position vector $\vec{r}'$. For example, if function $f$ is the function of position vector $\vec{r}$, that is $f=f(\vec{r})$  we can generally write:

\begin{equation}
\nabla f(\vec{r}) \neq 0 \hspace{2em}
\nabla' f(\vec{r}) = 0
\end{equation}

\noindent
On the other hand, if function $f$ is the function of position vector $\vec{r}'$, that is if $f=f(\vec{r}')$ we can write:

\begin{equation}
\nabla f(\vec{r}') = 0  \hspace{2em}
\nabla' f(\vec{r}') \neq 0
\end{equation}

\noindent
Furthermore, because variable $t'$ is independent of variables $x,y,z$ and $x',y',z'$ neither operator $\nabla$ nor $\nabla'$ acts on position vector $\vec{r}_s(t')$ and velocity vector $\vec{v}_s(t')$. Using these considerations we see that the following equations are correct:

\begin{equation}
\begin{array}{rcr}
\nabla\cdot\vec{r}_s(t')=0 & & \nabla\cdot\vec{v}_s(t')=0 \\
\nabla'\cdot\vec{r}_s(t')=0 & & \nabla'\cdot\vec{v}_s(t')=0
\end{array}
\end{equation}

\noindent
However, both operators $\nabla$ and $\nabla'$ act on Green's function $G(\vec{r},t;\vec{r}',t')$ given by equation (\ref{eq_fund_wave}). In fact, one can easily verify that the following equations hold:

\begin{equation}
\begin{array}{rcr}
\nabla G(\vec{r},t;\vec{r}',t') = -\nabla' G(\vec{r},t;\vec{r}',t')
&
&
\nabla^2 G(\vec{r},t;\vec{r}',t') = \nabla'^2 G(\vec{r},t;\vec{r}',t')
\\
\displaystyle
\frac{\partial}{\partial t} 
G(\vec{r},t;\vec{r}',t')
= -\frac{\partial}{\partial t'}G(\vec{r},t;\vec{r}',t') 
&
&
\displaystyle
\frac{\partial^2}{\partial t^2}
G(\vec{r},t;\vec{r}',t')
=
\frac{\partial^2}{\partial t'^2}
G(\vec{r},t;\vec{r}',t')
\end{array}
\end{equation}

\subsection{Generalized Helmholtz decomposition theorem}\label{sec_gen_helmholtz}
\noindent
To start deriving generalized Helmholtz decomposition theorem for vector functions of space and time we first consider inhomogeneous transient wave equation:

\begin{equation}
\label{apx_gen_helm_1}
\nabla^2 G(\vec{r},t;\vec{r}',t')
-
\frac{1}{c^2}
\frac{\partial^2}{\partial t^2}
G(\vec{r},t;\vec{r}',t')
=
-\delta(\vec{r}-\vec{r}')\delta(t-t')
\end{equation}

\noindent
where $G(\vec{r},\vec{t};\vec{r}',t')$ is the function called fundamental solution or Green's function and  $\delta$ is Dirac's delta function. The Green's function for inhomogeneous wave equation is well known and it represents an outgoing diverging spherical wave:

\begin{equation}
\label{apx_gen_helm_2}
G(\vec{r},t;\vec{r}',t')
=
\frac{
\delta\left(
t-t'
- \frac{\left|\vec{r}-\vec{r}'\right|}{c}
\right)
}{
4\pi\left|\vec{r}-\vec{r}'\right|
}
\end{equation}

\noindent
Let us now suppose that vector field $\vec{F}$ is the function of both space $\vec{r}$ and time $t$, i.e. $\vec{F}=\vec{F}(\vec{r},t)$. Using sifting property of Dirac delta function we can write vector function $\vec{F}(\vec{r},t)$ as the volume integral over infinite volume $\mathbb{R}^3$ and over all the time $\mathbb{R}$ as:

\begin{equation}
\label{apx_gen_helm_3}
\vec{F}(\vec{r},t)
=
\int_{\mathbb{R}}{dt'
\int_{\mathbb{R}^3}{
\vec{F}(\vec{r}',t')
\delta(\vec{r}-\vec{r}')
\delta(t-t')
dV'
}
}
\end{equation}

\noindent
where differential volume element $dV'$ is $dV'=dx'dy'dz'$.
We now replace $\delta(\vec{r}-\vec{r}')\delta(t-t')$ in equation above with left hand side of equation (\ref{apx_gen_helm_1}) to obtain:

\begin{equation}
\label{apx_gen_helm_4}
\vec{F}(\vec{r},t)
=
-
\int_{\mathbb{R}}{dt'
\int_{\mathbb{R}^3}{
\vec{F}(\vec{r}',t')
\left[
\nabla^2 G(\vec{r},t;\vec{r}',t')
-
\frac{1}{c^2}
\frac{\partial^2}{\partial t^2}
G(\vec{r},t;\vec{r}',t')
\right]
dV'
}
}
\end{equation}

\noindent
From the discussion presented in section \ref{sec:apx_identities_prelim} of this appendix, we know that D'Alambert operator $\nabla^2-\frac{1}{c^2}\frac{\partial^2}{\partial t^2}$ does not act on variables $x'$, $y'$,$z'$ and $t'$ nor does it act on vector function $\vec{F}(\vec{r}',t')$. Hence, we can write the D'Alambert operator $\nabla^2-\frac{1}{c^2}\frac{\partial^2}{\partial t^2}$ in front of the integral:

\begin{equation}
\label{apx_gen_helm_5}
\vec{F}(\vec{r},t)
=
-
\left(
\nabla^2
-
\frac{1}{c^2}
\frac{\partial^2}{\partial t^2}
\right)
\int_{\mathbb{R}}{dt'
\int_{\mathbb{R}^3}{
\vec{F}(\vec{r}',t')
G(\vec{r},t;\vec{r}',t')
dV'
}
}
\end{equation}

\noindent
Using standard vector calculus identity $\nabla\times\nabla\times\vec{P}=\nabla(\nabla\cdot\vec{P})-\nabla^2\vec{P}$ we can rewrite equation (\ref{apx_gen_helm_5}) as:

\begin{align}
\label{apx_gen_helm_6}
\vec{F}(\vec{r},t)
=&
\nabla\times\nabla\times
\int_{\mathbb{R}}{dt'
\int_{\mathbb{R}^3}{
\vec{F}(\vec{r}',t')
G(\vec{r},t;\vec{r}',t')
dV'
}
}
\\
&
-
\nabla
\left(
\nabla\cdot
\int_{\mathbb{R}}{dt'
\int_{\mathbb{R}^3}{
\vec{F}(\vec{r}',t')
G(\vec{r},t;\vec{r}',t')
dV'
}
}
\right)
\nonumber
\\
&
+
\frac{1}{c^2}
\frac{\partial^2}{\partial t^2}
\int_{\mathbb{R}}{dt'
\int_{\mathbb{R}^3}{
\vec{F}(\vec{r}',t')
G(\vec{r},t;\vec{r}',t')
dV'
}
}
\nonumber
\end{align}

\noindent
Because operators $\nabla$ and $\frac{\partial}{\partial t}$ do not act on variables $x'$, $y'$, $z'$ and $t'$ we can move operator $\nabla$ and partial derivative $\frac{\partial}{\partial t}$ under right hand side integrals in equation (\ref{apx_gen_helm_6}). Then using standard vector calculus identities $\nabla\times(\psi\vec{P})=\nabla\psi\times\vec{P}+\psi\nabla\times\vec{P}$ and $\nabla\cdot(\psi\vec{P})=\nabla\psi\cdot\vec{P}+\psi\nabla\cdot\vec{P}$, and noting that $\nabla\times\vec{F}(\vec{r}',t')=0$ and $\nabla\cdot\vec{F}(\vec{r}',t')=0$ we can rewrite equation (\ref{apx_gen_helm_6}) as:

\begin{align}
\label{apx_gen_helm_7}
\vec{F}(\vec{r},t)
=&
\nabla\times
\int_{\mathbb{R}}{dt'
\int_{\mathbb{R}^3}{
\nabla
G(\vec{r},t;\vec{r}',t')
\times
\vec{F}(\vec{r}',t')
dV'
}
}
\\
&
-
\nabla
\int_{\mathbb{R}}{dt'
\int_{\mathbb{R}^3}{
\vec{F}(\vec{r}',t')
\cdot
\nabla
G(\vec{r},t;\vec{r}',t')
dV'
}
}
\nonumber
\\
&
+
\frac{1}{c^2}
\frac{\partial}{\partial t}
\int_{\mathbb{R}}{dt'
\int_{\mathbb{R}^3}{
\vec{F}(\vec{r}',t')
\frac{\partial}{\partial t}
G(\vec{r},t;\vec{r}',t')
dV'
}
}
\nonumber
\end{align}

\noindent
We now use identities $\nabla G(\vec{r},\vec{t};\vec{r}',t') = -\nabla' G(\vec{r},\vec{t};\vec{r}',t')$ and $\frac{\partial}{\partial t}G(\vec{r},\vec{t};\vec{r}',t')=-\frac{\partial}{\partial t'}G(\vec{r},\vec{t};\vec{r}',t')$ to rewrite the right hand side integrals in equation (\ref{apx_gen_helm_7}) as:

\begin{align}
\label{apx_gen_helm_8}
\vec{F}(\vec{r},t)
=&
-
\nabla\times
\int_{\mathbb{R}}{dt'
\int_{\mathbb{R}^3}{
\nabla'
G(\vec{r},t;\vec{r}',t')
\times
\vec{F}(\vec{r}',t')
dV'
}
}
\\
&
+
\nabla
\int_{\mathbb{R}}{dt'
\int_{\mathbb{R}^3}{
\vec{F}(\vec{r}',t')
\cdot
\nabla'
G(\vec{r},t;\vec{r}',t')
dV'
}
}
\nonumber
\\
&
-
\frac{1}{c^2}
\frac{\partial}{\partial t}
\int_{\mathbb{R}}{dt'
\int_{\mathbb{R}^3}{
\vec{F}(\vec{r}',t')
\frac{\partial}{\partial t'}
G(\vec{r},t;\vec{r}',t')
dV'
}
}
\nonumber
\end{align}

\noindent
Using vector calculus identity $\nabla\times(\psi\vec{P})=\nabla\psi\times\vec{P}+\psi\nabla\times\vec{P}$ and the the form of divergence theorem $\int_{V}{\nabla\times\vec{P} dV}=\oint_{\partial V}{\vec{P}\times d\vec{S}}$ we rewrite the first right hand side integral over $\mathbb{R}^3$ as:

\begin{align}
\label{apx_gen_helm_9}
\int_{\mathbb{R}^3}{
\nabla'
G(\vec{r},t;\vec{r}',t')
\times
\vec{F}(\vec{r}',t')
dV'
}
=&
\oint_{\partial \mathbb{R}^3}{
G(\vec{r},t;\vec{r}',t')
\vec{F}(\vec{r}',t')
\times
d\vec{S}'
}
\\
&-
\int_{\mathbb{R}^3}{
G(\vec{r},t;\vec{r}',t')
\nabla'
\times
\vec{F}(\vec{r}',t')
dV'
}
\nonumber
\end{align}

\noindent
Note that the surface $\partial\mathbb{R}^3$ is an infinite surface that bounds an infinite volume $\mathbb{R}^3$. Furthermore, for the surface integral in the equation above, position vector $\vec{r'}$ is located on infinite surface $\partial\mathbb{R}^3$, i.e. $\vec{r}'\in\partial\mathbb{R}^3$. Hence, if vector function $\vec{F}(\vec{r}',t')$ decreases faster than $1/\left|\vec{r}-\vec{r}'\right|$ as $\left|\vec{r}-\vec{r}'\right|\rightarrow\infty$ the surface integral in equation (\ref{apx_gen_helm_9}) vanishes. In that case, we can write:

\begin{equation}
\label{apx_gen_helm_10}
\int_{\mathbb{R}^3}{
\nabla'
G(\vec{r},t;\vec{r}',t')
\times
\vec{F}(\vec{r}',t')
dV'
}
=
-
\int_{\mathbb{R}^3}{
G(\vec{r},t;\vec{r}',t')
\nabla'
\times
\vec{F}(\vec{r}',t')
dV'
}
\end{equation}

\noindent
Using similar considerations, vector calculus identity $\nabla\cdot(\psi\vec{P})=\nabla\psi\cdot\vec{P}+\psi\nabla\cdot\vec{P}$ and standard divergence theorem $\int_{V}{\nabla\cdot\vec{P}dV}=\oint_{\partial V}{\vec{P}\cdot d\vec{S}}$ it is obtained that:

\begin{equation}
\label{apx_gen_helm_11}
\int_{\mathbb{R}^3}{
\vec{F}(\vec{r}',t')
\cdot
\nabla'
G(\vec{r},t;\vec{r}',t')
dV'
}
=
-
\int_{\mathbb{R}^3}{
G(\vec{r},t;\vec{r},t')
\nabla'
\cdot
\vec{F}(\vec{r}',t')
dV'
}
\end{equation}

\noindent
To treat the last integral on the right hand side of equation (\ref{apx_gen_helm_8}) we  use the following identity:

\begin{equation}
\label{apx_gen_helm_12}
\vec{F}(\vec{r}',t')
\frac{\partial}{\partial t'}
G(\vec{r},t;\vec{r},t')
=
\frac{\partial}{\partial t'}
\left(
\vec{F}(\vec{r}',t')
G(\vec{r},t;\vec{r},t')
\right)
-
G(\vec{r},t;\vec{r},t')
\frac{\partial}{\partial t'}
\vec{F}(\vec{r}',t')
\end{equation}

\noindent
Using the identity above and noting that $t'$ is independent of $x'$, $y'$ and $z'$ we can rewrite the last right hand side integral of equation
(\ref{apx_gen_helm_8}) as:

\begin{align}
\label{apx_gen_helm_14}
\int_{\mathbb{R}}{dt'
\int_{\mathbb{R}^3}{
\vec{F}(\vec{r}',t')
\frac{\partial}{\partial t'}
G(\vec{r},t;\vec{r}',t')
dV'
}
}
=& 
\int_{\mathbb{R}}{dt'
\frac{\partial}{\partial t'}
\int_{\mathbb{R}^3}{
\vec{F}(\vec{r}',t')
G(\vec{r},t;\vec{r}',t')
dV'
}
}
\\
&-
\int_{\mathbb{R}}{dt'
\int_{\mathbb{R}^3}{
G(\vec{r},t;\vec{r}',t')
\frac{\partial}{\partial t'}
\vec{F}(\vec{r}',t')
dV'
}
}
\nonumber
\end{align}

\noindent
By integrating over $t'$, it can be shown that the first right hand side integral in equation (\ref{apx_gen_helm_14}) vanishes :

\begin{align}
\label{apx_gen_helm_15}
\int_{\mathbb{R}}{dt'
\frac{\partial}{\partial t'}
\int_{\mathbb{R}^3}{
\vec{F}(\vec{r}',t')
G(\vec{r},t;\vec{r}',t')
dV'
}
}
=\left[
\int_{\mathbb{R}^3}{
\vec{F}(\vec{r}',t')
G(\vec{r},t;\vec{r}',t')
dV'
}
\right]_{t'\rightarrow-\infty}^{t'\rightarrow\infty}
\end{align}

\noindent
If $t'\rightarrow\pm\infty$, and if $t$ is finite, then from equation (\ref{apx_gen_helm_2}) follows that $G(\vec{r},t;\vec{r}',t') = 0$, thus, the right hand side of equation (\ref{apx_gen_helm_15}) is equal to zero.  Inserting this result into equation (\ref{apx_gen_helm_14}) yields:

\begin{equation}
\label{apx_gen_helm_16}
\int_{\mathbb{R}}{dt'
\int_{\mathbb{R}^3}{
\vec{F}(\vec{r}',t')
\frac{\partial}{\partial t'}
G(\vec{r},t;\vec{r}',t')
dV'
}
}
=
-
\int_{\mathbb{R}}{dt'
\int_{\mathbb{R}^3}{
G(\vec{r},t;\vec{r}',t')
\frac{\partial}{\partial t'}
\vec{F}(\vec{r}',t')
dV'
}
}
\end{equation}

\noindent
By inserting equations (\ref{apx_gen_helm_10}), (\ref{apx_gen_helm_11}) and (\ref{apx_gen_helm_16}) into equation (\ref{apx_gen_helm_8}) we obtain the generalized Helmholtz theorem for vector functions of space and time:

\begin{align}
\label{apx_gen_helm_17}
\vec{F}(\vec{r},t)
=
&-\nabla \int_{\mathbb{R}}{
dt'
\int_{\mathbb{R}^3}{
\bigg(
\nabla'\cdot\vec{F}(\vec{r}',t')
\bigg)
G(\vec{r},t;\vec{r}',t')
dV'
}
}
\\
&+
\frac{1}{c^2}
\frac{\partial}{\partial t}
 \int_{\mathbb{R}}{
dt'
\int_{\mathbb{R}^3}{
\left(
\frac{\partial}{\partial t'}\vec{F}(\vec{r}',t')
\right)
G(\vec{r},t;\vec{r}',t')
dV'
}
}
\nonumber
\\
&+\nabla\times \int_{\mathbb{R}}{
dt'
\int_{\mathbb{R}^3}{
\bigg(
\nabla'\times\vec{F}(\vec{r}',t')
\bigg)
G(\vec{r},t;\vec{r}',t')
dV'
}
}
\nonumber
\end{align}

\noindent
The theorem is valid for functions $\vec{F}(\vec{r}',t')$ that decrease faster than  $1/\left|\vec{r}-\vec{r}'\right|$ as $\left|\vec{r}-\vec{r}'\right|\rightarrow\infty$.

\renewcommand{\theequation}{\Alph{section}.\arabic{equation}}
\section{Novel vector calculus identities}\label{sec:apx_vecalc_identities}
In this appendix we prove two novel vector calculus identities, without which, it would be very difficult, perhaps even not possible, to derive Maxwell's equations from Coulomb's law. These two novel vector calculus identities are given by the following two equations:

\begin{align}
\label{apx_new_identities_1}
\int_{V}{\psi\nabla^2\vec{P}dV}
&=
\oint_{\partial V}
{
\psi\left(d\vec{S}\cdot\nabla\right)\vec{P}
}
-
\int_{V}{
\left(\nabla\psi\cdot\nabla\right)\vec{P}dV
}
\\
\label{apx_new_identities_2}
\int_{V}{\vec{P}\nabla^2\psi dV}
&=
\oint_{\partial V}{
\vec{P}\left(\nabla\psi\cdot d\vec{S}\right)
}
-
\int_{V}{\left(\nabla\psi\cdot\nabla\right)\vec{P} dV}
\end{align}

\noindent
where $\vec{P}$ is differentiable vector field, $\psi$ is differentiable scalar function,
volume $V\subset\mathbb{R}^3$ is simply connected volume, $\partial V$ is the bounding surface of volume $V$ and $d\vec{S}$ is differential surface element of $\partial V$ such that $d\vec{S}=\vec{n} dS$. Vector $\vec{n}$ is an outward unit normal to the surface $\partial V$. In Cartesian coordinate system the product $\psi\nabla^2\vec{P}$ can be written in terms of Cartesian components as:

\begin{equation}
\label{apx_new_identities_3}
\psi\nabla^2\vec{P}
=
\hat{\vec{x}} \psi\nabla^2 P_x 
+
\hat{\vec{y}} \psi\nabla^2 P_y 
+
\hat{\vec{z}} \psi\nabla^2 P_z 
\end{equation}

\noindent
where $P_x$, $P_y$ and $P_z$ are Cartesian components of vector $\vec{P}$ and vectors $\vec{\hat{x}}$, $\vec{\hat{y}}$ and $\vec{\hat{z}}$ are Cartesian unit basis vectors. Using standard vector calculus identity $\nabla\cdot f \vec{T} = \nabla f \cdot \vec{T} + f \nabla\cdot\vec{T}$, valid for some scalar function $f$ and for some vector function $\vec{T}$, we can rewrite equation (\ref{apx_new_identities_3}) as:

\begin{align}
\label{apx_new_identities_4}
\psi\nabla^2\vec{P} &=
\hat{\vec{x}} \nabla\cdot\left(\psi \nabla P_x\right)
-
\hat{\vec{x}} \nabla\psi\cdot\nabla P_x
\\
&
+
\hat{\vec{y}} \nabla\cdot\left(\psi \nabla P_y\right)
-
\hat{\vec{y}} \nabla\psi\cdot\nabla P_y
\nonumber
\\
&
+
\hat{\vec{z}} \nabla\cdot\left(\psi \nabla P_z\right)
-
\hat{\vec{z}} \nabla\psi\cdot\nabla P_z
\nonumber
\end{align}

\noindent
To proceed, we now expand the identity $\left(\nabla\psi\cdot\nabla\right)\vec{P}$ in terms of its Cartesian components as:

\begin{align}
\label{apx_new_identities_5}
\left(\nabla\psi\cdot\nabla\right)\vec{P}
&=
\left(
\frac{\partial\psi}{\partial x}\frac{\partial}{\partial x}
+
\frac{\partial\psi}{\partial y}\frac{\partial}{\partial y}
+
\frac{\partial\psi}{\partial z}\frac{\partial}{\partial z}
\right)
\left(
\vec{\hat{x}}P_x + \vec{\hat{y}}P_y + \vec{\hat{z}}P_z  
\right)
\\
&=
\vec{\hat{x}} \nabla\psi\cdot\nabla P_x
+
\vec{\hat{y}} \nabla\psi\cdot\nabla P_y
+
\vec{\hat{z}} \nabla\psi\cdot\nabla P_z
\nonumber
\end{align}

\noindent
Inserting equation (\ref{apx_new_identities_5}) into (\ref{apx_new_identities_4}) it is obtained that:

\begin{equation}
\label{apx_new_identities_6}
\psi\nabla^2\vec{P} =
\hat{\vec{x}} \nabla\cdot\left(\psi \nabla P_x\right)
+
\hat{\vec{y}} \nabla\cdot\left(\psi \nabla P_y\right)
+
\hat{\vec{z}} \nabla\cdot\left(\psi \nabla P_z\right)
-
\left(\nabla\psi\cdot\nabla\right)\vec{P}
\end{equation}

\noindent
We now integrate equation (\ref{apx_new_identities_6}) over volume $V$
and apply the divergence theorem $\int_{V}{\nabla\cdot\vec{T}dV}=\oint_{\partial V}{\vec{T}\cdot d\vec{S}}$ to obtain:

\begin{align}
\label{apx_new_identities_7}
\int_{V}{\psi\nabla^2\vec{P}dV} =&
\hat{\vec{x}}\oint_{\partial V}{\psi\nabla P_x\cdot d\vec{S}}
+
\hat{\vec{y}}\oint_{\partial V}{\psi\nabla P_y\cdot d\vec{S}}
+
\hat{\vec{z}}\oint_{\partial V}{\psi\nabla P_z\cdot d\vec{S}}
\\
&
-
\int_{V}{\left(\nabla\psi\cdot\nabla\right)\vec{P}dV}
\nonumber
\end{align} 

\noindent
The first three right hand side terms of equation (\ref{apx_new_identities_7}) can be rewritten as:

\begin{equation}
\label{apx_new_identities_8}
\hat{\vec{x}}\oint_{\partial V}{\psi\nabla P_x\cdot d\vec{S}}
+
\hat{\vec{y}}\oint_{\partial V}{\psi\nabla P_y\cdot d\vec{S}}
+
\hat{\vec{z}}\oint_{\partial V}{\psi\nabla P_z\cdot d\vec{S}}
=
\oint_{\partial V}{\psi\left(d\vec{S}\cdot\nabla\right)\vec{P}}
\end{equation}

\noindent
Inserting equation (\ref{apx_new_identities_8}) into (\ref{apx_new_identities_7}) yields:

\begin{equation}
\label{apx_new_identities_9}
\int_{V}{\psi\nabla^2\vec{P}dV}
=
\oint_{\partial V}
{
\psi\left(d\vec{S}\cdot\nabla\right)\vec{P}
}
-
\int_{V}{
\left(\nabla\psi\cdot\nabla\right)\vec{P}dV
}
\end{equation}

\noindent
which we intended to prove. To prove equation (\ref{apx_new_identities_2}) we rewrite $\vec{P}\nabla^2\psi$ in terms of Cartesian components as:

\begin{equation}
\label{apx_new_identities_10}
\vec{P}\nabla^2\psi
=
\vec{\hat{x}} P_x \nabla^2 \psi
+
\vec{\hat{y}} P_y \nabla^2 \psi
+
\vec{\hat{z}} P_z \nabla^2 \psi
\end{equation}

\noindent
By using standard differential calculus identity $f\nabla^2 h = \nabla\cdot(f\nabla h)-\nabla f\cdot\nabla h$, where $f$ and $h$ are differentiable scalar functions, equation (\ref{apx_new_identities_10}) can be written as:

\begin{align}
\label{apx_new_identities_11}
\vec{P}\nabla^2\psi
=&
\vec{\hat{x}} \nabla\cdot\left(P_x\nabla\psi\right)
-\vec{\hat{x}} \nabla P_x \cdot \nabla\psi +
\\
&\vec{\hat{y}} \nabla\cdot\left(P_y\nabla\psi\right)
-\vec{\hat{y}} \nabla P_y \cdot \nabla\psi +
\nonumber
\\
&
\vec{\hat{z}} \nabla\cdot\left(P_z\nabla\psi\right)
-\vec{\hat{z}} \nabla P_z \cdot \nabla\psi
\nonumber
\end{align}

\noindent
Inserting equation (\ref{apx_new_identities_5}) into equation (\ref{apx_new_identities_11}) yields:

\begin{equation}
\label{apx_new_identities_12}
\vec{P}\nabla^2\psi
=
\vec{\hat{x}} \nabla\cdot\left(P_x\nabla\psi\right)
+
\vec{\hat{y}} \nabla\cdot\left(P_y\nabla\psi\right)
+
\vec{\hat{z}} \nabla\cdot\left(P_z\nabla\psi\right)
-
\left(\nabla\psi\cdot\nabla\right)\vec{P}
\end{equation}

\noindent
Integrating equation (\ref{apx_new_identities_12}) over volume $V$
and applying divergence theorem  $\int_{V}{\nabla\cdot\vec{T}dV}=\oint_{\partial V}{\vec{T}\cdot d\vec{S}}$ it is obtained that:

\begin{equation}
\label{apx_new_identities_14}
\int_{V}{\vec{P}\nabla^2\psi dV}
=
\oint_{\partial V}{
\vec{P}\left(\nabla\psi\cdot d\vec{S}\right)
}
-
\int_{V}{\left(\nabla\psi\cdot\nabla\right)\vec{P} dV}
\end{equation}

\noindent
which we intended to prove.

\renewcommand{\theequation}{\Alph{section}.\arabic{subsection}.\arabic{equation}}
\section{Derivation of auxiliary mathematical identities}\label{sec:apx_identities}
In this appendix we derive auxiliary mathematical identities that we find useful for the derivation of Maxwell equations from Coulomb's law.

\subsection{Derivation of equation (\ref{eq_coulomb_d3})}\label{sec:apx_identities_d1}

\noindent
Equation (\ref{eq_utility_1}) allows us to rewrite 
the time derivative in the second right hand side integral in equation (\ref{eq_coulomb_d2})  as:

\begin{equation}
\label{eq_apx1_1}
\frac{\partial}{\partial t'}\frac{q_s}{4 \pi \epsilon}
\frac{\vec{r}'-\vec{r}_s(t')}
{\left|\vec{r}'-\vec{r}_s(t')\right|^3}
=
-
\frac{q_s}{4 \pi \epsilon}
\frac{\partial}{\partial t'}
\nabla' \frac{1}{\left|\vec{r}'-\vec{r}_s(t')\right|}
\end{equation}

\noindent
Because coordinates $x'$, $y'$ and $z'$ are independent of time $t'$  we can swap operator $\nabla'$ and time derivative with respect to time $t'$ as:

\begin{equation}
\label{eq_apx1_2}
\frac{\partial}{\partial t'}\frac{q_s}{4 \pi \epsilon}
\frac{\vec{r}'-\vec{r}_s(t')}
{\left|\vec{r}'-\vec{r}_s(t')\right|^3}
=
-
\frac{q_s}{4 \pi \epsilon}
\nabla' 
\frac{\partial}{\partial t'}
\frac{1}{\left|\vec{r}'-\vec{r}_s(t')\right|}
\end{equation}

\noindent
Furthermore, because coordinates $x'$, $y'$ and $z'$ are independent of time $t'$,
the time derivative of $\vec{r}'$ is equal to zero $\frac{\partial \vec{r}'}{\partial t'}=0$. The inner time derivative in the equation (\ref{eq_apx1_2}) can now be written as:

\begin{equation}
\label{eq_apx1_3}
\frac{\partial}{\partial t'}
\frac{1}{\left|\vec{r}'-\vec{r}_s(t')\right|}
=
\frac{
\vec{v}_s(t')\cdot\left(\vec{r}'-\vec{r}_s(t')\right)
}{
\left|\vec{r}'-\vec{r}_s(t')\right|^3
}
=
-
\nabla'
\cdot 
\frac{\vec{v}_s(t')}{\left|\vec{r}'-\vec{r}_s(t')\right|}
\end{equation}

\noindent
where $\vec{v}_s(t')$ is the velocity of the source charge $q_s$ at time $t'$ given by equation (\ref{eq_prelim_4}).
Inserting equation (\ref{eq_apx1_3}) into equation (\ref{eq_apx1_2}) yields:

\begin{equation}
\label{eq_apx1_5}
\frac{\partial}{\partial t'}\frac{q_s}{4 \pi \epsilon}
\frac{\vec{r}'-\vec{r}_s(t')}
{\left|\vec{r}'-\vec{r}_s(t')\right|^3}
=
\frac{q_s}{4 \pi \epsilon}
\nabla'
\left(
\nabla'
\cdot
\frac{\vec{v}_s(t')}{\left|\vec{r}'-\vec{r}_s(t')\right|}
\right)
\end{equation}

\noindent
To proceed with derivation, we now make use of standard vector calculus identity $\nabla\times\nabla\times\vec{P}=\nabla\left(\nabla\cdot\vec{P}\right)-\nabla^2\vec{P}$, valid for any differentiable vector function $\vec{P}$. This identity allows us to rewrite the  equation (\ref{eq_apx1_5}) as:

\begin{equation}
\label{eq_apx1_6}
\frac{\partial}{\partial t'}\frac{q_s}{4 \pi \epsilon}
\frac{\vec{r}'-\vec{r}_s(t')}
{\left|\vec{r}'-\vec{r}_s(t')\right|^3}
=
\frac{q_s}{4 \pi \epsilon}
\nabla'
\times
\nabla'
\times
\frac{\vec{v}_s(t')}{\left|\vec{r}'-\vec{r}_s(t')\right|}
+
\frac{q_s}{4 \pi \epsilon}
\nabla'^2
\frac{\vec{v}_s(t')}{\left|\vec{r}'-\vec{r}_s(t')\right|}
\end{equation}

\noindent
Since Laplacian operator $\nabla'^2$ does not have effect on  velocity vector $\vec{v}_s(t')$ the last right hand side term in equation (\ref{eq_apx1_6}) can be written using 3D Dirac's delta function as:

\begin{equation}
\label{eq_apx1_7}
\frac{q_s}{4 \pi \epsilon}
\nabla'^2
\frac{\vec{v}_s(t')}{\left|\vec{r}'-\vec{r}_s(t')\right|}
=
\frac{q_s}{\epsilon} \vec{v}_s(t')
\nabla'^2
\frac{1}{4\pi}
\frac{1}{\left|\vec{r}'-\vec{r}_s(t')\right|}
=
- \frac{q_s}{\epsilon} \vec{v}_s(t') \delta\left(\vec{r}'-\vec{r}_s(t')\right)
\end{equation}

\noindent
Hence, replacing the last right hand side term of equation (\ref{eq_apx1_6}) with equation (\ref{eq_apx1_7}) yields:

\begin{equation}
\label{eq_apx1_8}
\frac{\partial}{\partial t'}\frac{q_s}{4 \pi \epsilon}
\frac{\vec{r}'-\vec{r}_s(t')}
{\left|\vec{r}'-\vec{r}_s(t')\right|^3}
=
\frac{q_s}{4 \pi \epsilon}
\nabla'
\times
\nabla'
\times
\frac{\vec{v}_s(t')}{\left|\vec{r}'-\vec{r}_s(t')\right|}
- \frac{q_s}{\epsilon} \vec{v}_s(t') \delta\left(\vec{r}'-\vec{r}_s(t')\right)
\end{equation}

\noindent
which proves equation (\ref{eq_coulomb_d3}).

\subsection{Derivation of equation (\ref{eq_coulomb_d10})}\label{sec:apx_identities_d2}
To derive equation (\ref{eq_coulomb_d10}) we make use of standard vector calculus identity $\nabla\times\left(\psi\vec{P}\right)=\nabla\psi\times\vec{P}+\psi\nabla\times\vec{P}$, where $\psi$ is a scalar function and $\vec{P}$ is a vector function, to rewrite the integrand in the last right hand side term of equation (\ref{eq_coulomb_d9}) as:

\begin{eqnarray}
\label{eq_apx2_1}
\lefteqn{
\left[
\nabla'
\times
\nabla'
\times
\frac{\vec{v}_s(t')}{\left|\vec{r}'-\vec{r}_s(t')\right|}
\right]
G(\vec{r},t;\vec{r}',t')
=
}
\\
&&=
\nabla'\times
\left[
\left(
\nabla'
\times
\frac{\vec{v}_s(t')}{\left|\vec{r}'-\vec{r}_s(t')\right|}
\right)
G(\vec{r},t;\vec{r}',t')
\right]
-
\nabla'G(\vec{r},t;\vec{r}',t')
\times
\left[
\nabla'
\times
\frac{\vec{v}_s(t')}{\left|\vec{r}'-\vec{r}_s(t')\right|}
\right]
\nonumber
\end{eqnarray}

\noindent
Integrating equation (\ref{eq_apx2_1}) with respect to variables $x'$, $y'$, $z'$ and $t'$, and making use of a standard form of divergence theorem $\int_{V}{\nabla\times\vec{P}dV}=\oint_{\partial V}{d\vec{S}\times\vec{P}}$ it is obtained that:

\begin{eqnarray}
\label{eq_apx2_2}
\lefteqn{
\int_{\mathbb{R}}{
dt'
\int_{\mathbb{R}^3}
{
\left[
\nabla'
\times
\nabla'
\times
\frac{\vec{v}_s(t')}{\left|\vec{r}'-\vec{r}_s(t')\right|}
\right]
G(\vec{r},t;\vec{r}',t')
dV'
}
}
=
}
\\
&=&
\int_{\mathbb{R}}{
dt'
\oint_{\partial\mathbb{R}^3}
{
d\vec{S}'
\times
\left(
\nabla'
\times
\frac{\vec{v}_s(t')}{\left|\vec{r}'-\vec{r}_s(t')\right|}
\right)
G(\vec{r},t;\vec{r}',t')
}
}
\nonumber
\\
&&
-
\int_{\mathbb{R}}
{
dt'
\int_{\mathbb{R}^3}
{
\nabla'G(\vec{r},t;\vec{r}',t')
\times
\left[
\nabla'
\times
\frac{\vec{v}_s(t')}{\left|\vec{r}'-\vec{r}_s(t')\right|}
\right]
dV'
}
}
\nonumber
\end{eqnarray}

\noindent
where $dV=dx'dy'dz'$, $\partial\mathbb{R}^3$ is an infinite surface that bounds $\mathbb{R}^3$, and $d\vec{S}$ is differential surface element of the surface $\partial\mathbb{R}^3$. Because $\partial\mathbb{R}^3$ is an infinite surface, the first right hand side integral vanishes. To see this, we can use standard vector identity $\nabla\times(\psi\vec{P})=\nabla\psi\times\vec{P}+\psi\nabla\times\vec{P}$ and using $\nabla'\times\vec{v}_s(t')=0$ to rewrite the first term in the first right hand side integrand as:

\begin{equation}
\label{eq_apx2_3}
\nabla'
\times
\frac{\vec{v}_s(t')}{\left|\vec{r}'-\vec{r}_s(t')\right|}
=
-\frac{\vec{r}'-\vec{r}_s(t')}{\left|\vec{r}'-\vec{r}_s(t')\right|^3}
\times\vec{v}_s(t')
\end{equation}

\noindent
Because $\vec{r}'\in\partial\mathbb{R}^3$ this means that $\left|\vec{r'}\right|\rightarrow\infty$. Provided that charge $q_s$ is moving with finite velocity $\vec{v}_s(t')$ it is clear that right hand side term of equation (\ref{eq_apx2_3}) vanishes as $\left|\vec{r'}\right|\rightarrow\infty$. Thus, we can now write equation (\ref{eq_apx2_2}) as:

\begin{eqnarray}
\label{eq_apx2_4}
\lefteqn{
\int_{\mathbb{R}}{
dt'
\int_{\mathbb{R}^3}
{
\left[
\nabla'
\times
\nabla'
\times
\frac{\vec{v}_s(t')}{\left|\vec{r}'-\vec{r}_s(t')\right|}
\right]
G(\vec{r},t;\vec{r}',t')
dV'
}
}
=
}
\\
&=&
-
\int_{\mathbb{R}}
{
dt'
\int_{\mathbb{R}^3}
{
\nabla'G(\vec{r},t;\vec{r}',t')
\times
\left[
\nabla'
\times
\frac{\vec{v}_s(t')}{\left|\vec{r}'-\vec{r}_s(t')\right|}
\right]
dV'
}
}
\nonumber
\end{eqnarray}

\noindent
There is another useful property of Green's function $G(\vec{r},t;\vec{r}',t')$ which enables us to proceed with the derivation of equation (\ref{eq_coulomb_d10}). This property can be written as follows:

\begin{equation}
\label{eq_apx2_5}
\nabla' G(\vec{r},t;\vec{r}',t')
=
-\nabla G(\vec{r},t;\vec{r}',t')
\end{equation}

\noindent
Using this property and standard vector calculus identity $\nabla\times(\psi\vec{P})=\nabla\psi\times\vec{P}+\psi\nabla\times\vec{P}$ allows us to rewrite the integrand in equation (\ref{eq_apx2_2}) as:

\begin{eqnarray}
\label{eq_apx2_6}
\lefteqn{
\nabla'G(\vec{r},t;\vec{r}',t')
\times
\left[
\nabla'
\times
\frac{\vec{v}_s(t')}{\left|\vec{r}'-\vec{r}_s(t')\right|}
\right]
=
}
\\
&&=
-\nabla G(\vec{r},t;\vec{r}',t')
\times
\left[
\nabla'
\times
\frac{\vec{v}_s(t')}{\left|\vec{r}'-\vec{r}_s(t')\right|}
\right]
=
\nonumber
\\
&&=
-\nabla
\times
\left[
G(\vec{r},t;\vec{r}',t')
\nabla'
\times
\frac{\vec{v}_s(t')}{\left|\vec{r}'-\vec{r}_s(t')\right|}
\right]
\nonumber
\end{eqnarray}

\noindent
Equation above is valid because operator $\nabla$ does not act on velocity vector $\vec{v}_s(t')$, nor does it act on position vectors $\vec{r}'$ and $\vec{r}_s(t')$. It only acts on Green's function $G(\vec{r},t;\vec{r}',t')$ because it is a function of position vector $\vec{r}$. Inserting equation (\ref{eq_apx2_6}) into equation (\ref{eq_apx2_4}) yields:

\begin{eqnarray}
\label{eq_apx2_7}
\lefteqn{
\int_{\mathbb{R}}{
dt'
\int_{\mathbb{R}^3}
{
\left[
\nabla'
\times
\nabla'
\times
\frac{\vec{v}_s(t')}{\left|\vec{r}'-\vec{r}_s(t')\right|}
\right]
G(\vec{r},t;\vec{r}',t')
dV'
}
}
=
}
\\
&=&
\int_{\mathbb{R}}
{
dt'
\int_{\mathbb{R}^3}
{
\nabla
\times
\left[
G(\vec{r},t;\vec{r}',t')
\nabla'
\times
\frac{\vec{v}_s(t')}{\left|\vec{r}'-\vec{r}_s(t')\right|}
\right]
dV'
}
}
\nonumber
\end{eqnarray}

\noindent
Because differential volume element is $dV'=dx'dy'dz'$ and because operator $\nabla$ does not act on variables $x'$, $y'$, $z'$ and $t'$ we can write operator $\nabla$ in front of the integral:

\begin{eqnarray}
\label{eq_apx2_8}
\lefteqn{
\int_{\mathbb{R}}{
dt'
\int_{\mathbb{R}^3}
{
\left[
\nabla'
\times
\nabla'
\times
\frac{\vec{v}_s(t')}{\left|\vec{r}'-\vec{r}_s(t')\right|}
\right]
G(\vec{r},t;\vec{r}',t')
dV'
}
}
=
}
\\
&=&
\nabla\times
\int_{\mathbb{R}}
{
dt'
\int_{\mathbb{R}^3}
{
G(\vec{r},t;\vec{r}',t')
\nabla'
\times
\frac{\vec{v}_s(t')}{\left|\vec{r}'-\vec{r}_s(t')\right|}
dV'
}
}
\nonumber
\end{eqnarray}

\noindent
Using the same trick again, that is,
by using standard vector calculus identity $\nabla\times(\psi\vec{P})=\nabla\psi\times\vec{P}+\psi\nabla\times\vec{P}$, using $\nabla' G(\vec{r},t;\vec{r}',t')
=
-\nabla G(\vec{r},t;\vec{r}',t')$ and noting that operator $\nabla$ does not act on variables $x'$, $y'$ and $z'$ we can rewrite the integrand in equation (\ref{eq_apx2_8}) as:

\begin{eqnarray}
\label{eq_apx2_9}
\lefteqn{
G(\vec{r},t;\vec{r}',t')
\nabla'
\times
\frac{\vec{v}_s(t')}{\left|\vec{r}'-\vec{r}_s(t')\right|}
=
}
\\
&=&
\nabla'\times\left[
G(\vec{r},t;\vec{r}',t')
\frac{\vec{v}_s(t')}{\left|\vec{r}'-\vec{r}_s(t')\right|}
\right]
+
\nabla
\times
\left[
G(\vec{r},t;\vec{r}',t')
\frac{\vec{v}_s(t')}{\left|\vec{r}'-\vec{r}_s(t')\right|}
\right]
\nonumber
\end{eqnarray}

\noindent
Then, by inserting equation (\ref{eq_apx2_9}) into equation (\ref{eq_apx2_8}) and using a form of standard divergence theorem $\int_{V}{\nabla\times\vec{P}dV}=\oint_{\partial V}{d\vec{S}\times\vec{P}}$ we obtain that:

\begin{eqnarray}
\label{eq_apx2_10}
\lefteqn{
\int_{\mathbb{R}}{
dt'
\int_{\mathbb{R}^3}
{
\left[
\nabla'
\times
\nabla'
\times
\frac{\vec{v}_s(t')}{\left|\vec{r}'-\vec{r}_s(t')\right|}
\right]
G(\vec{r},t;\vec{r}',t')
dV'
}
}
=
}
\\
&=&
\nabla\times
\int_{\mathbb{R}}
{
dt'
\oint_{\partial\mathbb{R}^3}
{
d\vec{S}'\times
\frac{\vec{v}_s(t')}{\left|\vec{r}'-\vec{r}_s(t')\right|}
}
} +
\nonumber
\\
&&
\nabla\times\
\int_{\mathbb{R}}
{
dt'
\int_{\partial\mathbb{R}^3}
{
\nabla
\times
\left[
G(\vec{r},t;\vec{r}',t')
\frac{\vec{v}_s(t')}{\left|\vec{r}'-\vec{r}_s(t')\right|}
\right]
dV'
}
} 
\nonumber
\end{eqnarray}

\noindent
Because surface $\partial\mathbb{R}^3$ is an infinite surface the magnitude of position vector $\vec{r}'$ is infinite, hence the surface integral in the first right hand side term of equation (\ref{eq_apx2_10}) vanishes. Furthermore, because operator $\nabla$ does not act on variables $x'$, $y'$, $z'$ and $t'$ we can write operator $\nabla$ in front of the second right hand side space-time integral. Hence, equation (\ref{eq_apx2_10}) can be written as:

\begin{eqnarray}
\label{eq_apx2_11}
\lefteqn{
\int_{\mathbb{R}}{
dt'
\int_{\mathbb{R}^3}
{
\left[
\nabla'
\times
\nabla'
\times
\frac{\vec{v}_s(t')}{\left|\vec{r}'-\vec{r}_s(t')\right|}
\right]
G(\vec{r},t;\vec{r}',t')
dV'
}
}
=
}
\\
&=&
\nabla\times\nabla\times
\int_{\mathbb{R}}
{
dt'
\int_{\partial\mathbb{R}^3}
{
G(\vec{r},t;\vec{r}',t')
\frac{\vec{v}_s(t')}{\left|\vec{r}'-\vec{r}_s(t')\right|}
dV'
}
} 
\nonumber
\end{eqnarray}

\noindent
thus, proving the equation (\ref{eq_coulomb_d10}).

\subsection{Derivation of equation (\ref{eq_coulomb_d18})}\label{sec:apx_identities_d3}
Using the mathematical identity $\nabla^2 G(\vec{r},t;\vec{r}',t)=\nabla'^2 G(\vec{r},t;\vec{r}',t)$, valid for Green's function $G(\vec{r},t;\vec{r}',t)$, one can rewrite the left hand side integral in equation (\ref{eq_coulomb_d18}) 
as:

\begin{eqnarray}
\label{apx_identities_d3_1}
\lefteqn{
\int_{\mathbb{R}}{
dt'
\int_{\mathbb{R}^3}{
\frac{q_s}{4 \pi \epsilon}
\frac{\vec{v}_s(t')}{\left|\vec{r}'-\vec{r}_s(t')\right|}
\nabla^2
G(\vec{r},t;\vec{r}',t')
dV'
}
}
=
}
\\
&=&
\int_{\mathbb{R}}{
dt'
\int_{\mathbb{R}^3}{
\frac{q_s}{4 \pi \epsilon}
\frac{\vec{v}_s(t')}{\left|\vec{r}'-\vec{r}_s(t')\right|}
\nabla'^2
G(\vec{r},t;\vec{r}',t')
dV'
}
}
\nonumber
\end{eqnarray}

\noindent
In Appendix \ref{sec:apx_vecalc_identities}, we have derived two novel vector calculus identities. Subtracting vector identity (\ref{apx_new_identities_1}) from vector identity (\ref{apx_new_identities_2}) yields:

\begin{equation}
\label{apx_identities_d3_4}
\int_{V}{\vec{P}\nabla^2\psi dV}
=
\int_{V}{\psi\nabla^2\vec{P}dV}
+
\oint_{\partial V}{
\vec{P}\left(\nabla\psi\cdot d\vec{S}\right)
}
-
\oint_{\partial V}
{
\psi\left(d\vec{S}\cdot\nabla\right)\vec{P}
}
\end{equation}

\noindent
Using vector identity (\ref{apx_identities_d3_4}) we can rewrite equation (\ref{apx_identities_d3_1}) as:

\begin{eqnarray}
\label{apx_identities_d3_5}
\lefteqn{
\int_{\mathbb{R}}{
dt'
\int_{\mathbb{R}^3}{
\frac{q_s}{4 \pi \epsilon}
\frac{\vec{v}_s(t')}{\left|\vec{r}'-\vec{r}_s(t')\right|}
\nabla^2
G(\vec{r},t;\vec{r}',t')
dV'
}
}
=
}
\\
&=&
\int_{\mathbb{R}}{
dt'
\int_{\mathbb{R}^3}{
\frac{q_s}{4 \pi \epsilon}
G(\vec{r},t;\vec{r}',t')
\nabla'^2
\frac{\vec{v}_s(t')}{\left|\vec{r}'-\vec{r}_s(t')\right|}
dV'
}
}
+
\nonumber
\\
&&
\int_{\mathbb{R}}
{
dt'
\oint_{\partial\mathbb{R}^3}
{
\frac{q_s}{4 \pi \epsilon}
\frac{\vec{v}_s(t')}{\left|\vec{r}'-\vec{r}_s(t')\right|}
\left(
\nabla'
G(\vec{r},t;\vec{r}',t')
\cdot
d\vec{S}'
\right)
}
}
-
\nonumber
\\
&&
\int_{\mathbb{R}}
{
dt'
\oint_{\partial\mathbb{R}^3}
{
G(\vec{r},t;\vec{r}',t')
\left(
d\vec{S}'\cdot\nabla'
\right)
\frac{q_s}{4 \pi \epsilon}
\frac{\vec{v}_s(t')}{\left|\vec{r}'-\vec{r}_s(t')\right|}
}
}
\nonumber
\end{eqnarray}

\noindent
Because surface $\partial\mathbb{R}^3$ is an infinite surface the magnitude of position vector $\vec{r}'\in\partial\mathbb{R}^3$ has an infinite magnitude, $\left|\vec{r}'\right|\rightarrow\infty$. In that case, both right hand side surface integrals over surface $\partial\mathbb{R}^3$ vanish in equation (\ref{apx_identities_d3_5}). Hence, equation (\ref{apx_identities_d3_5}) becomes:

\begin{eqnarray}
\label{apx_identities_d3_6}
\lefteqn{
\int_{\mathbb{R}}{
dt'
\int_{\mathbb{R}^3}{
\frac{q_s}{4 \pi \epsilon}
\frac{\vec{v}_s(t')}{\left|\vec{r}'-\vec{r}_s(t')\right|}
\nabla^2
G(\vec{r},t;\vec{r}',t')
dV'
}
}
=
}
\\
&=&
\int_{\mathbb{R}}{
dt'
\int_{\mathbb{R}^3}{
\frac{q_s}{\epsilon}
G(\vec{r},t;\vec{r}',t')
\nabla'^2
\frac{\vec{v}_s(t')}{4\pi\left|\vec{r}'-\vec{r}_s(t')\right|}
dV'
}
}
\nonumber
\end{eqnarray}

\noindent
The operator $\nabla'$ does not affect vector $\vec{v}_s(t')$ which is a function of variable $t'$. Thus, we can rewrite the Laplacian in the equation above as:

\begin{equation}
\label{apx_identities_d3_7}
\nabla'^2
\frac{\vec{v}_s(t')}{4\pi\left|\vec{r}'-\vec{r}_s(t')\right|}
=
\vec{v}_s(t')
\nabla'^2
\frac{1}{4\pi\left|\vec{r}'-\vec{r}_s(t')\right|}
=
-\vec{v}_s(t')\delta\left(\vec{r}'-\vec{r}_s(t')\right)
\end{equation}

\noindent
Inserting equation (\ref{apx_identities_d3_7}) into right hand side of equation (\ref{apx_identities_d3_6}) yields:

\begin{eqnarray}
\label{apx_identities_d3_8}
\lefteqn{
\int_{\mathbb{R}}{
dt'
\int_{\mathbb{R}^3}{
\frac{q_s}{4 \pi \epsilon}
\frac{\vec{v}_s(t')}{\left|\vec{r}'-\vec{r}_s(t')\right|}
\nabla^2
G(\vec{r},t;\vec{r}',t')
dV'
}
}
=
}
\\
&&
-
\int_{\mathbb{R}}{
dt'
\int_{\mathbb{R}^3}{
\frac{q_s}{\epsilon}
\vec{v}_s(t')\delta\left(\vec{r}'-\vec{r}_s(t')\right)
G(\vec{r},t;\vec{r}',t')
dV'
}
}
\nonumber
\end{eqnarray}

\noindent
which proves equation (\ref{eq_coulomb_d18}).

\subsection{Derivation of equation (\ref{eq_coulomb_d39})}\label{sec:apx_identities_d4}
Equation (\ref{eq_coulomb_d39}) can  be derived from equation (\ref{eq_lorentz_7}) by taking the divergence of both sides of this equation to obtain:

\begin{equation}
\label{apx_identities_d4_1}
\nabla\cdot\vec{N}(\vec{r},t)
+
\nabla\cdot\vec{K}(\vec{r},t)
=
\nabla\cdot
\frac{ q_s}{4 \pi \epsilon}
\frac{\vec{r}-\vec{r}_s(t)}
{\left|\vec{r}-\vec{r}_s(t)\right|^3}
=
\frac{q_s}{\epsilon}\delta\left(\vec{r}-\vec{r}_s(t)\right)
\end{equation}

\noindent
To find $\nabla\cdot\vec{K}(\vec{r},t)$ we can write operator $\nabla$ under the right hand side integral of equation (\ref{eq_lorentz_5}) and then apply identity $\nabla G(\vec{r},t;\vec{r}',t)=-\nabla' G(\vec{r},t;\vec{r}',t)$ to obtain:

\begin{equation}
\label{apx_identities_d4_2}
\nabla\cdot
\vec{K}(\vec{r},t)
=-
\frac{1}{c^2}
\frac{\partial}{\partial t}
\int_{\mathbb{R}}{
dt'
\int_{\mathbb{R}^3}{
\frac{q_s}{4 \pi \epsilon}
\left[
\nabla'
\times
\nabla'
\times
\frac{\vec{v}_s(t')}{\left|\vec{r}'-\vec{r}_s(t')\right|}
\right]
\cdot
\nabla'
G(\vec{r},t;\vec{r}',t')
dV'
}
}
\end{equation}

\noindent
From here, using standard vector identity $\nabla\cdot(\psi\vec{P})=\nabla\psi\cdot\vec{P}+\psi\nabla\cdot\vec{P}$ and divergence theorem it is obtained that:

\begin{eqnarray}
\label{apx_identities_d4_3}
\lefteqn{
\nabla\cdot
\vec{K}(\vec{r},t)
=
}
\\
&=&-
\frac{1}{c^2}
\frac{\partial}{\partial t}
\int_{\mathbb{R}}{
dt'
\oint_{\partial\mathbb{R}^3}{
\frac{q_s}{4 \pi \epsilon}
\left[
\nabla'
\times
\nabla'
\times
\frac{\vec{v}_s(t')}{\left|\vec{r}'-\vec{r}_s(t')\right|}
\right]
G(\vec{r},t;\vec{r}',t')
\cdot d\vec{S}'
}
}
\nonumber
\\
&&+
\frac{1}{c^2}
\frac{\partial}{\partial t}
\int_{\mathbb{R}}{
dt'
\int_{\mathbb{R}^3}{
\frac{q_s}{4 \pi \epsilon}
\left[
\nabla'\cdot
\nabla'
\times
\nabla'
\times
\frac{\vec{v}_s(t')}{\left|\vec{r}'-\vec{r}_s(t')\right|}
\right]
G(\vec{r},t;\vec{r}',t')
dV'
}
}
\nonumber
\end{eqnarray}

\noindent
Clearly, the surface integral on the right hand side of equation (\ref{apx_identities_d4_3}) vanishes as $\left|\vec{r}'\right|\rightarrow 0$. Furthermore, because $\nabla\cdot\nabla\times\vec{P}=0$ the second right hand side term vanishes as well. Hence, we can write:

\begin{equation}
\label{apx_identities_d4_4}
\nabla\cdot\vec{K}(\vec{r},t)=0
\end{equation}

\noindent
Inserting equation (\ref{apx_identities_d4_4}) into equation (\ref{apx_identities_d4_1}) yields:

\begin{equation}
\nabla\cdot\vec{N}(\vec{r},t)
=
\frac{q_s}{\epsilon}\delta\left(\vec{r}-\vec{r}_s(t)\right)
\end{equation}

\noindent
thus, proving the equation (\ref{eq_coulomb_d39}).

\end{document}